\documentclass[a4paper]{article}

\usepackage[utf8]{inputenc}
\usepackage{amsmath}
\usepackage{graphicx}
\usepackage[colorinlistoftodos]{todonotes}
\usepackage{subfigure}
\usepackage{booktabs}
\usepackage{multirow}
\usepackage{siunitx}
\usepackage{url}
\usepackage{setspace}
\usepackage{multicol,lipsum}
\usepackage{fancyhdr}

\usepackage{color}

\fancypagestyle{plain}{
    \fancyhead{}
    \fancyhead[C]{ \textit{Research Report 6/2018} \\ National Laboratory for Scientific Computing (LNCC)   
    }
}
\title{\textbf{A Multilayer and Time-varying Structural Analysis of the Brazilian Air Transportation Network}}
\author{Bernardo Costa, Jo\~{a}o Victor Bechara, Klaus Wehmuth, Artur Ziviani\\
National Laboratory for Scientific Computing (LNCC)\\ Av. Get\'{u}lio Vargas, 333, Quitandinha\\
CEP 25651-075 -- Petr\'{o}polis, RJ -- Brazil\\
\url{{bantunes,joaovmb,klaus,ziviani}@lncc.br}
}

\date{}

\begin{document}

\maketitle

\noindent \textbf{\textit{Abstract.}}
\textit{This paper provides a multilayer and time-varying structural analysis of one air transportation network, having the Brazilian air transportation network as a case study. Using a single mathematical object called MultiAspect Graph~(MAG) for this analysis, the multi-layer perspective enables the unveiling of the particular strategies of each airline to both establish and adapt in a moment of crisis its specific flight network. Similarly, the time-varying perspective allows multi-scale analysis considering different time periods, and thus assessing the impact of the economic crisis on how the different airlines establish their routes as well as the flights that use these routes. Altogether, besides the multilayer and time-varying structural analysis of the Brazilian air transportation network, this paper also acts as a proof-of-concept for the MAG potential for the modeling and analysis of high-order networks.} 

\noindent \textbf{\textit{Resumo.}}
\textit{Este artigo prov\^{e} uma an\'{a}lise estrutural multicamada e variante no tempo da malha a\'{e}rea, tendo a malha a\'{e}rea brasileira como estudo de caso. Usando um \'{u}nico objeto matem\'{a}tico chamado 
Grafo MultiAspecto~(MAG) para essa an\'{a}lise, a perspectiva multicamada permite revelar as estrat\'{e}gias particulares de cada companhia a\'{e}rea tanto para estabelecer quanto adaptar em um momento de crise 
sua respectiva malha de voos. De forma similar, a perspectiva variante no tempo permite an\'{a}lise multi-escala considerando diferentes per\i{i}odos de tempo, e portanto permitindo a avalia\c{c}\~{a}o do impacto da 
crise econ\^{o}mica em como as diferentes companhias a\'{e}reas estabelecem suas rotas bem como os voos que usam essas rotas. Como um todo, al\'{e}m da an\'{a}lise estrutural multicamada e variante no tempo da malha a\'{e}rea brasileira, este artigo tamb\'{e}m atua como uma prova de conceito para o potencial de MAGs para a modelagem e an\'{a}lise de redes de alta ordem.}


\doublespacing

\section{Introduction}
\label{sec:intro}

The air transportation network is a fundamental component of the modern society with a large economic impact. Therefore, it's important to analyze the structure of the air transportation network available to the population and understand how the airline companies providing this service are prepared to match the demand changes over the years. In this context, the Brazilian domestic air transportation network constitutes a complex (i.e., non-trivial) network composed of a variety of airline companies with different strategies to define their own flight network while covering a large territory.
    
Recent analyses of the Brazilian domestic air transportation network focused on flight delay analysis~\cite{Sternberg2016} or aggregated structural analysis of the network~\cite{Couto2015}. Such recent structural analysis of the Brazilian domestic air transportation network~\cite{Couto2015}, as well as the global air transportation network~\cite{Verma2014, Wei2014}, were elaborated using aggregated network graphs, containing only information about airports (localities) and routes (links between airports). This type of analysis disregards the temporal information of scheduling of flights, making information about connecting flights (i.e. network paths) not reliable, since no time information is available to the flights. Therefore, in these network models, it is not possible to guarantee the time sequence of the connecting flights, since there is no temporal information in the route graph. The aggregated view of the air transportation network also conceals the structural subnetwork of each one of the airline companies, thus being oblivious to the particular strategies adopted by each company when defining their specific air transportation network.
       
In this paper, we analyze the Brazilian domestic air transportation network in such a way that all time scheduling information is preserved.  Furthermore, information identifying the company of each flight is also made available, so that the network may be studied as a whole, or using different perspectives such as views from independent airline companies, used routes, and performed flights. Previous work typically focused on a single aspect of the concerned flight network, such as the aggregated structure~\cite{Couto2015}, the multilayer perspective~\cite{Tsiotas2015}, or a property of the network such as flight delays~\cite{Sternberg2016}. Therefore, in contrast to previous work, we provide an integrated multilayer~\cite{Kivela2014} and time-varying~\cite{Holme2012,Wehmuth2015-dsaa} structural analysis, focusing on the core structural network under different perspectives and having the Brazilian air transportation network as a case study. 

All the presented results are obtained through the air transportation network represented by a single mathematical object, but still allowing a multi-layer time-varying view of the network. 
The multi-layer view unveils the strategy of each airline company when defining its air transportation network.
The adopted single mathematical object can also represent the air transportation network under study with temporal information on different time scales. Hence, it is possible to obtain the complete scheduling information of flights of all airlines considered in the study at different periods of time, thus allowing a comparative study on before and after the recent economical crisis that largely affected the Brazilian air transportation network. In particular, in combination with the multi-layer view, it is possible to reveal the differences in the strategy adopted by each company to couple with the recent economical crisis and the resulting change in the flight network.

This paper is organized as follows. Section \ref{method} describes our methodology. Section \ref{results} analyzes the results. Finally, in Section~\ref{conclusao:}, we conclude the paper. 

\section{Methodology}
\label{method}

In this section, we describe the methods and tools that we use in the model construction as well as to obtain the results presented in this article.

\subsection{MAG}

MultiAspect Graph (MAG)~\cite{Wehmuth2016} is a graph generalization that can represent high-order networks, such as multi-layer, time-varying, and multi-scale complex networks. In this context, an aspect is an independent feature of the complex networked system to be modeled, such as localities, layers, and time instants. One of the key characteristics of a MAG is to be isomorphic to a directed graph and a companion tuple~\cite{Wehmuth2017}. In this way, with a MAG, it is possible to apply the already available knowledge from graph theory for analyzing directed graphs directly into the MAG environment.

In particular, algorithms for application in MAGs can be developed based on algorithms already known for oriented graphs~\cite{Wehmuth2017}, facilitating the analysis of complex networks modeled by MAGs. In addition to the algorithms and operations typically used with directed graphs, a model based on MAGs allows making aggregations on aspects, what is called sub-determinations. The sub-determinations enable results expressed in function of a subset of the aspects present in the MAG, but always allowing the use of all the information found in the model to get the result.

In this work, we model the Brazilian air transportation network as a network composed of localities~(airports), layers~(corresponding to the individual air transportation network of each airline company), and time instants. However, the results obtained are expressed only regarding locations~(airports) and their links~(routes), implying the use of sub-determined algorithms. The achieved results take into account the time and layer structure of the model to respect the time sequence of available flights as well as the operating boundaries between different airlines.

\subsection{MAG Operations}

To extract the information used in the next sections we perform two operations on the MAG representing the Brazilian air transportation network. Namely, the operations we describe are the Sub-MAG and the Sub-Determination, which we detail in the following subsections.

\subsubsection{Sub-MAG}
\label{subsubsec:sub-mag}

A sub-MAG is a part of a MAG that is also a MAG. This operation is used to extract only a fraction of the MAG for a detailed analysis. For example, to analyze the MAG corresponding to the air transportation network of a single airline company, we performed the Sub-MAG operation with that company. The result is a MAG containing only information corresponding to the chosen company. The sub-MAG operation is directly related to the well-known (induced) sub-graph operation on directed graphs.

\subsubsection{Sub-determination}

Sub-determination is the operation that enables the removal of an aspect of a MAG for an aggregated view of the MAG. For example, to analyze the connections between airports, regardless of the airline company, we can perform a sub-determination of the MAG by the airline company,  extracting a MAG without this aspect. The sub-determination is similar to the known aggregation operation on multilayer and time-varying graphs, extending the aggregation concept so that it can by applied to any proper subset of the aspects of a MAG. Therefore, given a MAG with $p$ aspects, $2^p -2$ distinct sub-determinations are possible~\cite{Wehmuth2016,Wehmuth2017}.

\subsection{K-Core}

\begin{figure}[!ht]
  \begin{center}
  \includegraphics[width=0.7\textwidth]{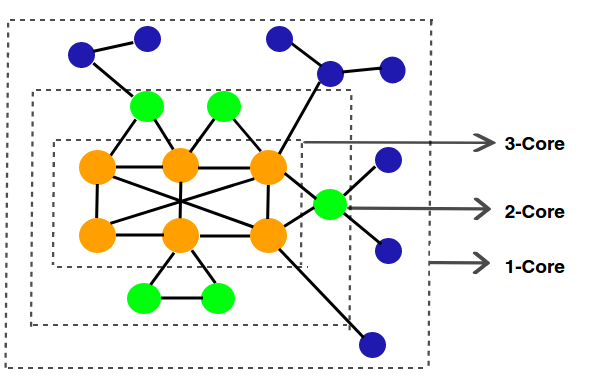}
  \caption{Illustrative example of the K-Core algorithm.}
  \label{fig:k-core}
  \end{center}
\end{figure}

In our analysis, we use the K-Core algorithm~\cite{SEIDMAN1983} to identify the network core. The K-Core of a network is obtained through a continuous decomposition of the network that removes all vertices with connectivity lower than the value of $k$ and their respective connections.  For example, eliminating all vertices of degree~1, the result is a core with vertices with at least degree~2. The resulting graph in this case is the 2-core. Increasing the value of $k$, the final round of the algorithm happens when it is impossible to eliminate more vertices, in other words, the network with (k+1)-core is empty. In other words, in this case, the maximum K-Core is achieved. Note that  for two cores to be considered equal they must be composed of the same vertices and edges, i.e., they must have the vertices linked by the same topological structure.
Figure~\ref{fig:k-core} represents an illustrative example of how the K-Core algorithm works. 

By adopting the K-Core algorithm to analyze the air transportation network, the intent in this paper is to focus on a structural analysis of the most relevant vertices of the Brazilian air transportation network~(i.e., those that compose the core of the network) under different perspectives as discussed in Section~\ref{results}.

\section{Results}
\label{results}
This section presents the adopted datasets corresponding to the Brazilian air transportation network in two different periods of time and the structural analysis of this network taking into account different viewpoints, such as time-varying, multilayer, and multi-scale perspectives. 

\subsection{Model construction}
To represent the Brazilian domestic air transportation network, we use two flight schedules: one from June 3, 2015, and the other from May 13, 2016, both published by the National Civil Aviation Agency~(ANAC) on its website. These schedule tables contain information about domestic, international, postal, and cargo flights covering a period of one week.  We extract from these tables the information about domestic commercial passenger flights, the target of this paper, on both periods~(2015 and 2016), separated by approximately 11~months.

The MAG-based model proposed in this paper to analyze the Brazilian air transportation network has four aspects~(airports, airlines, flight time instants, and the time period of the dataset). The first aspect contains all airports in Brazil that had at least one flight registered in the week each dataset refers to. There are 110 and 109 airports in the 2015 and in the 2016 dataset, respectively. The second aspect contains all the airline companies. In the 2015 dataset, there are seven airlines~(Azul, Avianca, Gol, Map, Passaredo, Tam, and Sete). In the 2016 dataset, there are eight airlines~(Azul, Avianca, Gol, Map, Passaredo, Tam, and Sete). We focus our structural
analysis on the four largest airline companies, namely Gol, TAM, Azul, and Avianca, as they together concentrate more than 90\% of the routes between airports and 95\% of the flights that use these routes in the Brazilian domestic air transportation network. 

To ease the modeling, we create two layers for each airline: The first layer represents the actual flights and the second layer represents the possible connections between flights, resulting in fourteen layers for the first dataset and in sixteen layers for the second one.
In the third aspect, there is information about all the flight times, in minutes, in a week. The forth aspect identifies the corresponding time period of the dataset: The value one is used for the 2015 dataset and the value two for the 2016 dataset.

\begin{figure}[ht]
\centering
	\subfigure[July 2015]{ 
    	\includegraphics[width=0.4\textwidth]{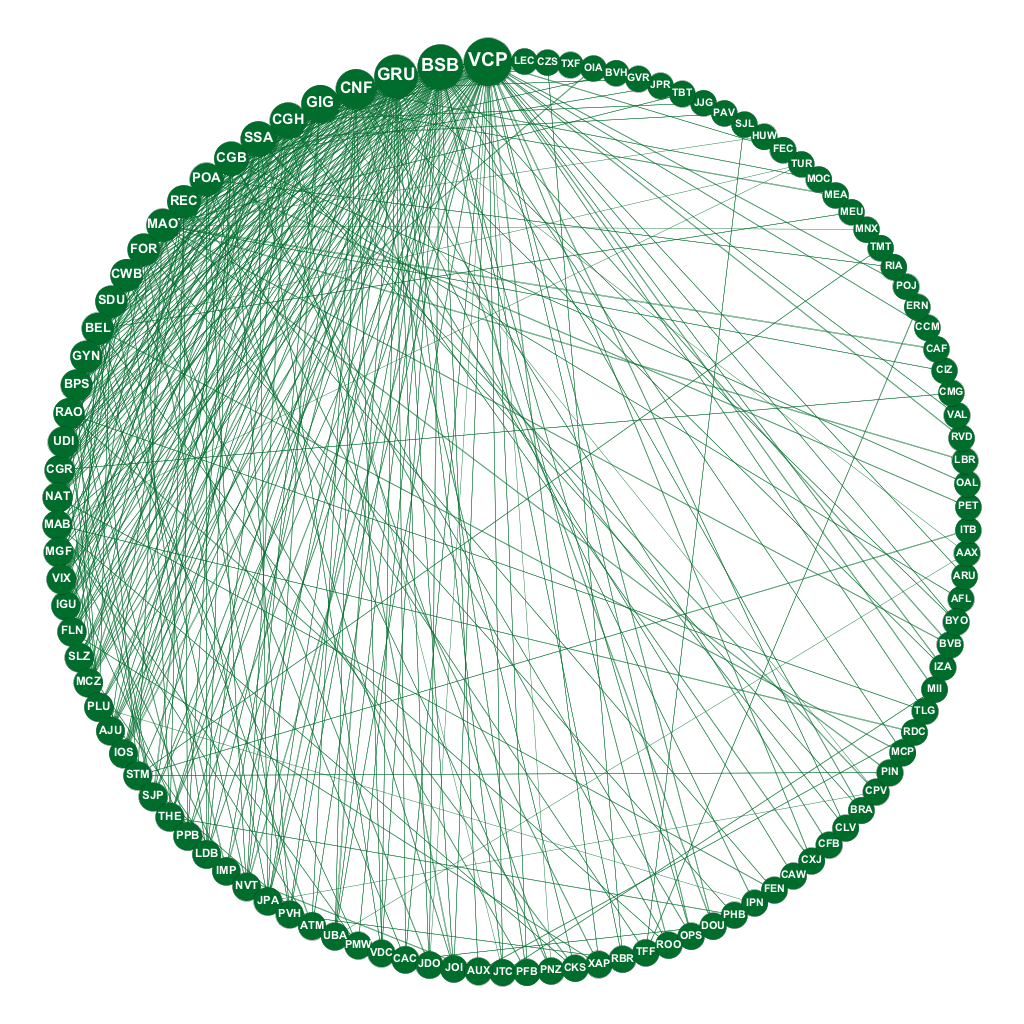}
      	\label{subfig:2015}
	   }
	\subfigure[May 2016]{
		\includegraphics[width=0.4\textwidth]{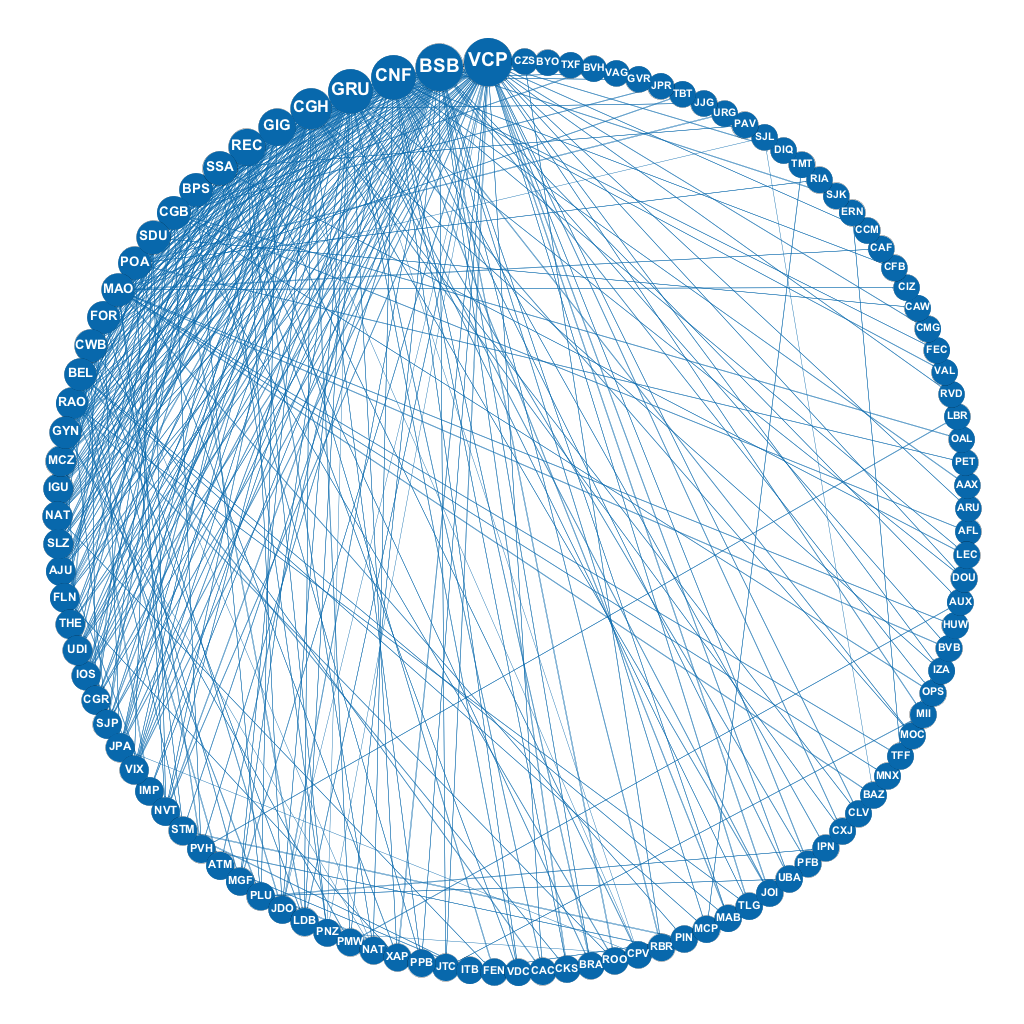}
	     \label{subfig:2016}
			  }
			  \caption{Brazilian domestic air transportation network.}
			  \label{fig:malhatotal}
			  \end{figure}
     
\subsection{Comparison of the Brazilian air transportation network: 2015 x 2016}
The severe economic crisis in Brazil in 2016 directly impacted several sectors of the economy, making them to adapt to the new reality. In this context, civil aviation is no exception, and there has been an adjustment on the part of the airline companies in their availability of flights, resulting in a change in the Brazilian domestic air transportation network from 2015 to 2016 that we start analyzing in this section. Figure~\ref{fig:malhatotal} shows the full Brazilian air transportation network for both the 2015 and 2016 datasets, sub-determined to show only the first MAG aspect, i.e., the locations~(airports). In this representation, a vertex is an airport and an edge is a route between two airports. Table~\ref{tab:num_air} shows the number of served airports by each airline in the Brazilian air transportation network in 2015 and 2016. There are some slight changes with some airlines serving less airports while others started serving more airports. imposing changes in the Brazilian air transportation network. We here analyze not only the changes in the number of served airports but most importantly we analyze how these airports are served in terms of routes between them and also in terms of the number of flights using these routes, and in particular we show that changes are in general more significant when taking those perspectives into account.

\begin{table}[t] 
\centering
\caption{Number of served airports in the Brazilian air transportation network.}
\label{tab:num_air}
\vspace{0.5cm}
\begin{tabular}{|c|c|c|c|}
\hline 
Company & \# Airports in 2015 & \# Airports in 2016 & Difference (\%)\\
\hline                               
Gol & 56 & 53 & -5\\
Azul & 100  & 94 & -6\\ 
TAM & 43 & 46  & +7\\
Avianca & 23 & 24 &  +4\\
\hline
\end{tabular}
\end{table}

Table~\ref{tab:num_rotas} shows a reduction of 15\% in the number of routes of the main airlines in the Brazilian air transportation network between June 2015 and May 2016. The three largest airlines in the country (Gol, Azul, and TAM) had significant reductions in their number of routes, specially Gol with a $25\%$ decrease in the number of routes in the time period. Nevertheless, Avianca increased the number of routes by 6\%, showing that it took advantage of the route reduction made by larger airlines to occupy some niches and expand. Therefore, the average growth of smaller companies in times of crisis is their strategy to partially fill the gaps in routes left by larger companies.

\begin{table}[t] 
\centering
\caption{Number of routes in the Brazilian air transportation network.}
\label{tab:num_rotas}
\vspace{0.5cm}
\begin{tabular}{|c|c|c|c|}
\hline 
Company & \# Routes in 2015 & \# Routes in 2016 & Difference (\%)\\
\hline                               
Gol & 324 & 242 & -25  \\
Azul & 454  & 377 & -17\\
TAM & 250 & 234 & -6\\
Avianca & 102 & 109 & +6\\
\hline
Total & 1130 & 962 & -15\\
\hline
\end{tabular}
\end{table}

Furthermore, Table~\ref{tab:num_voos} shows the difference in the number of flights between June 2015 and May 2016, indicating a 20\% decrease in the number of flights. In general, the changes in the number of flights follow a similar trend as in the change in the number of routes, but the intensity of the increase or reduction in the number of flights is even larger in the extreme cases. Considering the number of flights, for instance, Gol had a decrease of 37\%, whereas Avianca had an increase of 13\% in their number of flight, suggesting that the intensity that the airline companies use the route by offering more or less flights along a week in each route also represents a change in another dimension.

\begin{table}[t] 
\centering
\caption{Number of flights in the Brazilian air transportation network.}
\label{tab:num_voos}
\vspace{0.5cm}
\begin{tabular}{|c|c|c|c|}
\hline 
Company & \# Flights in 2015 & \# Flights in 2016 & Difference (\%)\\
\hline                               
Gol & 6644 & 4188 & -37\\
Azul & 5839  & 5075 & -13\\
TAM & 4814 & 4158 & -14\\
Avianca & 1386 & 1561 & +13\\
\hline
Total & 18683 & 14982 & -20\\
\hline
\end{tabular}
\end{table}

Additionally, \emph{codesharing} is regularly used to optimize occupation in flights on low volume routes. Considering the analyzed datasets, the number of cases of \emph{codesharing} went from 923 to 1234 flights per week, an increase of about $33\%$. This strategy virtually keeps a declared flight active in a per airline view, but it actually represents a reduction in the number of real flights on air.

\subsection{Digraph analysis: The route-based perspective}
\label{subsec:digraph}

A digraph is a simple directed graph.  In our context, a digraph represents the existing routes between the airports, disregarding how busy each route is~(i.e., how many daily flight compose the route). 
In this subsection, we analyze the Brazilian air transportation network with the digraph format. A previous paper uses this approach to analyze the Brazilian air transportation network~\cite{Couto2015}. With the digraph information, it is possible to quantify which airports have more connectivity in the air transportation network, for instance. 
In contrast to~\cite{Couto2015}, we apply the K-Core methodology to the Brazilian air transportation network. Moreover, we also apply such a methodology to the flight network of each airline company, allowing a number of different analyses concerning the structure of the network at its basic core structure. The main properties of the Brazilian air transportation network are also analyzed in two different time periods~(2015 and 2016). The resulting central core of the flight network of each airline company indicates the key airports for each company, thus unveiling the strategy of each company in defining its flight network. Therefore, the obtained information with the K-Core methodology allows the analysis of the concentration of the activities of each airline, a comparison of the strategies taken by the companies, and the identification of the hubs of the companies, for instance.

The analysis based on the K-Core methodology can be done in two ways: with representations in the digraph and in the multi-digraph format. In both, it is possible to separate the analysis for each company. In the digraph form, there is at most one edge between each an origin and a destination airport. This implies that the analysis considers only existing routes between airports, disregarding the number of flights using each route. In contrast, in the multi-digraph format, all flights are considered~(this case is analyzed in Section~\ref{subsec:multidigraph}). 
 
\begin{figure}[ht]
  \begin{center}
  \subfigure[2015]{
  \includegraphics[width=0.47\textwidth]{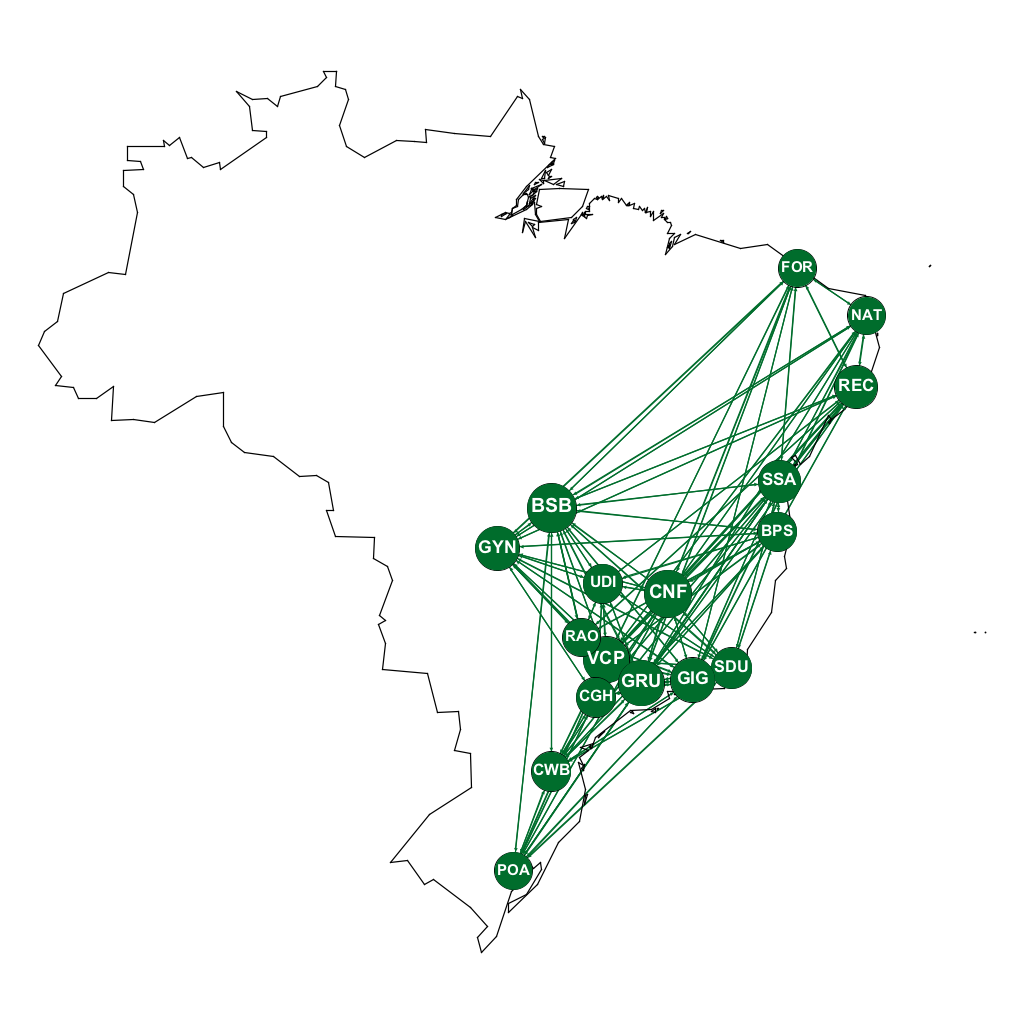}
  \label{fig:di-core-2015}
  }
  \subfigure[2016]{
  \includegraphics[width=0.47\textwidth]{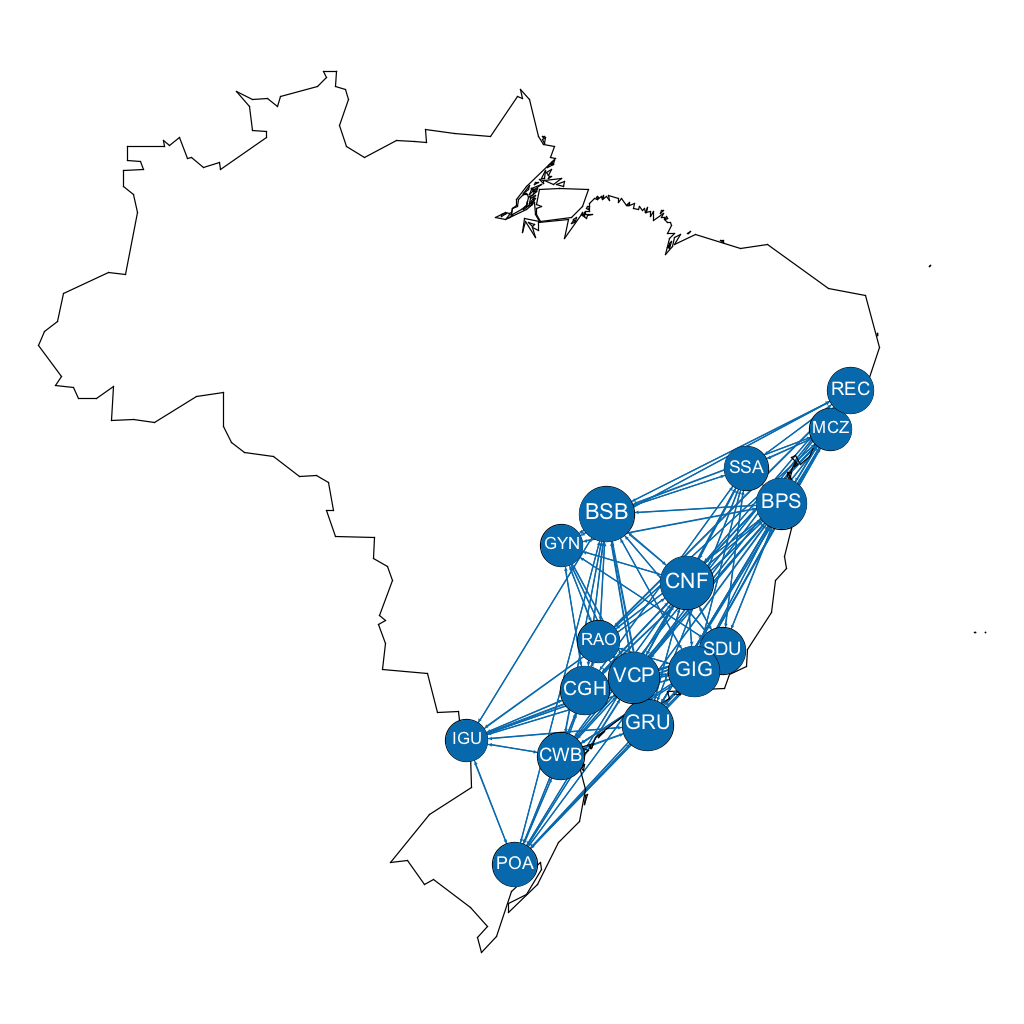}
    \label{fig:di-core-2016}
  }
  \caption{The geo-referenced K-Core of the Brazilian air transportation network with a digraph (route-based) perspective: (a) as of July 2015 for the maximum value of K=18 (17 airports and 196 routes in the maximum K-Core); and (b) as of May 2016 for the maximum K=18 (16 airports and 182 routes in the maximum K-Core). }
  \label{fig:di-core}
  \end{center}
\end{figure}

First, we analyze the connectivity of the maximum K-Core in the digraph corresponding to the whole Brazilian air transportation network, thus focusing on the structural analysis of the basic core of the flight network. Figure~\ref{fig:di-core-2015} and~\ref{fig:di-core-2016} show the geo-referenced K-Core of the digraph corresponding to the complete Brazilian air transportation network, as indicated by the 2015 and 2016 dataset, respectively. The 2015 core network results from the maximum K-Core with K=18, 17 airports, and~196 routes. Similarly, the 2016 core network results from the maximum K-Core with K=18, 16 airports, and~182 routes. As it could be expected, the south and southeast regions, which present the largest share in Brazil's economy, are also the regions that compose the central core of the air transportation network in the country, with a small a trend of even stronger concentration in 2016.

We now consider the connectivity of the maximum K-Core in the digraph corresponding to flight network of each of the four main airline companies in Brazil. Table~\ref{tab:core_dig} shows the maximum core number~$K$ as well as the number of airports and routes in the maximum K-Core of each of the main airline companies, both in June 2015 and May 2016. The maximum core number $K$ of all airlines remained the same between the two time periods, although the number of airports and routes in the maximum core of the main airline companies has been significantly reduced, except for Avianca, which kept the same core. This indicates that the main airlines have significantly reduced the number of key airports and routes among them in their core structures, although the same level of connectivity was kept. The exception here is Avianca that kept the same level on its structure (the same K-Core size) with only some slight changes in the structure of its core as discussed later. A more detailed analysis on how each of the main airlines adapted its core structure comes next.

\begin{table}  	
 \centering
 \caption{Comparison of maximum K-Cores in the digraph representation.}
 \label{tab:core_dig}
 \vspace{0.5cm}
  \begin{tabular}{|l|r|r|r|r|r|r|}
    \hline
    \multirow{2}{*}{Airlines} &
      \multicolumn{2}{c|}{Core (K)} &
      \multicolumn{2}{c|}{\# Airports} &
      \multicolumn{2}{c|}{\# Routes} \\
   & 2015 & 2016 & 2015 & 2016 & 2015 & 2016 \\
   \hline                               
    Gol & 12 & 12 & 13 & 8 & 106 & 50 \\
	Azul & 10 & 10 & 21 & 15 & 174 & 110 \\
	TAM & 10 & 10 & 13 & 10 & 92 & 64 \\
	Avianca & 8 & 8 & 8 & 8 & 42 & 42 \\
\hline 
  \end{tabular}
\end{table}

 \begin{figure}[hp]
\centering
	\subfigure[Gol's core on June 2015]{ 
    	\includegraphics[width=0.35\textwidth]{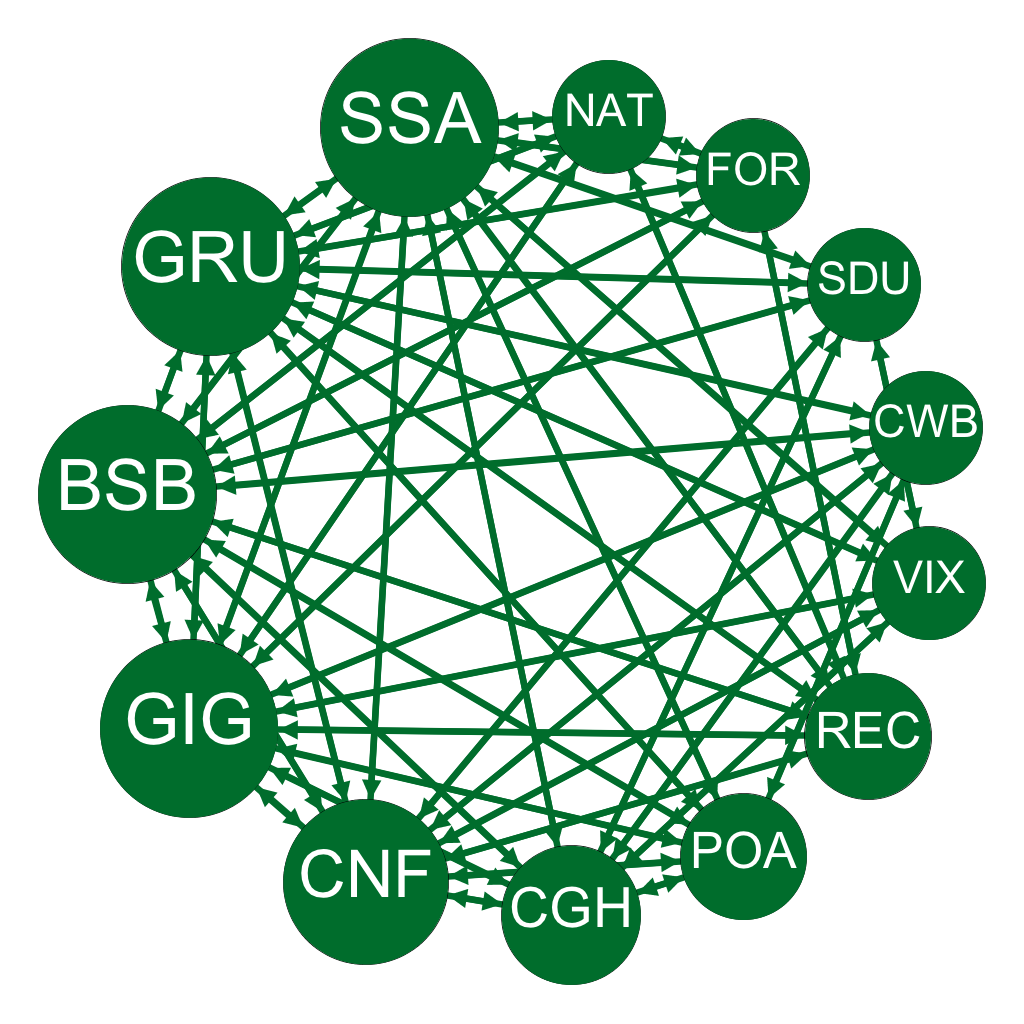}
      	\label{subfig:gol-core-2015}
	   }
	\subfigure[Gol's core on May 2016]{
		\includegraphics[width=0.35\textwidth]{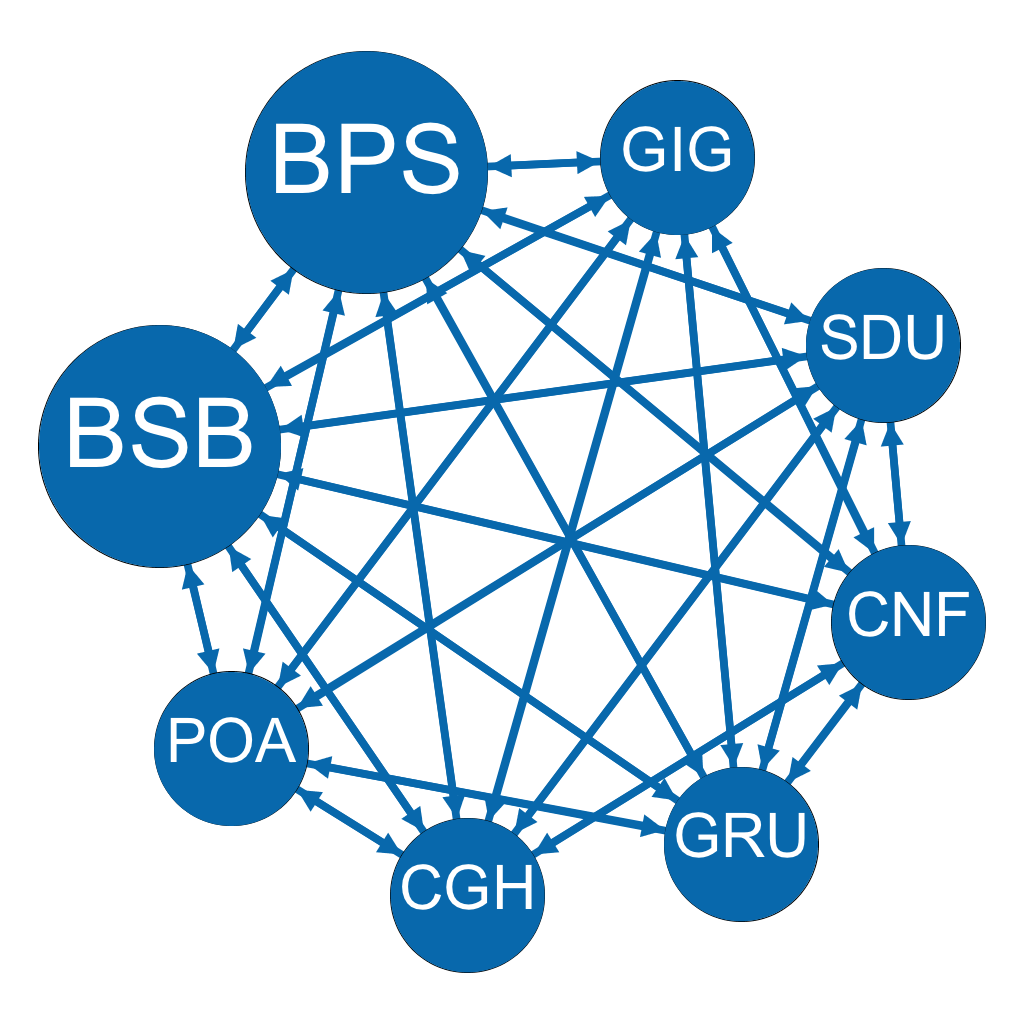}
	     \label{subfig:gol-core-2016}
			  }
	\subfigure[Azul's core on June 2015]{ 
    	\includegraphics[width=0.35\textwidth]{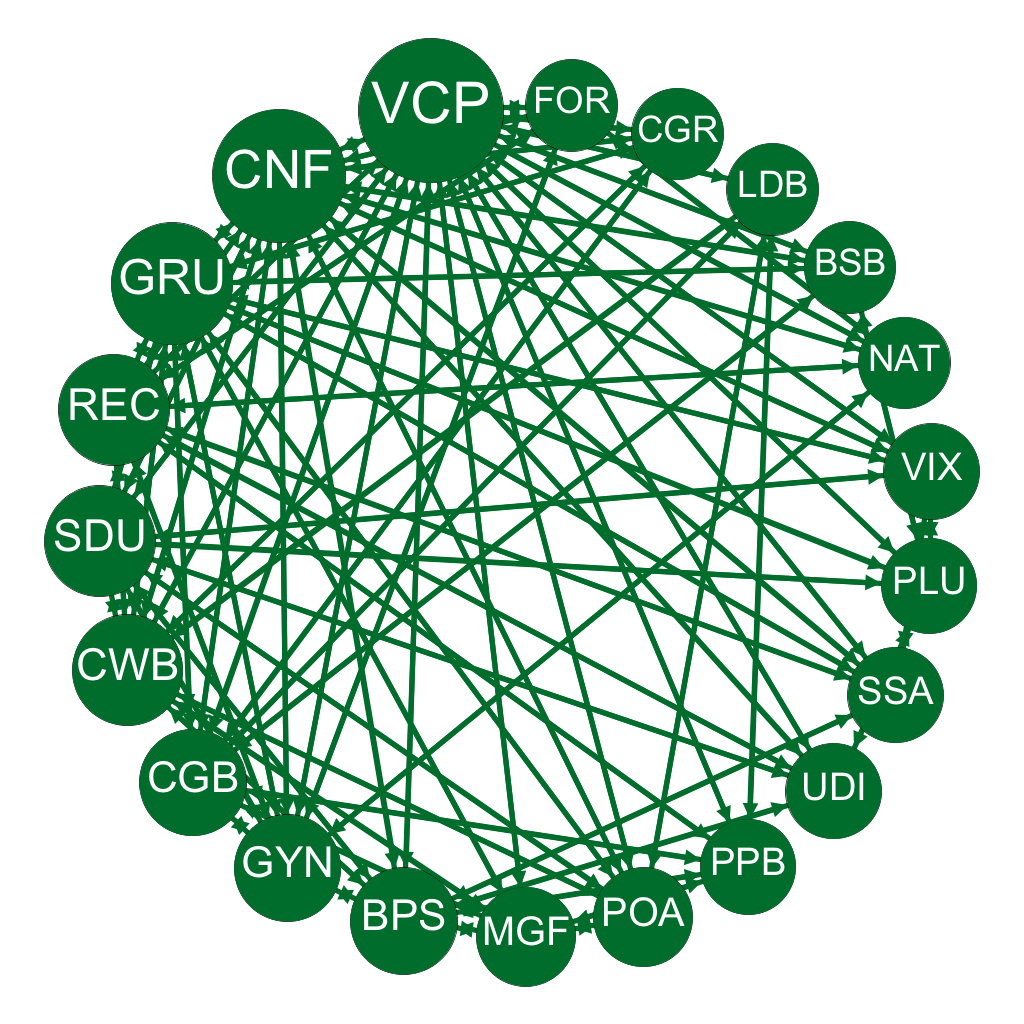}
      	\label{subfig:CDA1}
	   }
	\subfigure[Azul's core on May 2016]{
		\includegraphics[width=0.35\textwidth]{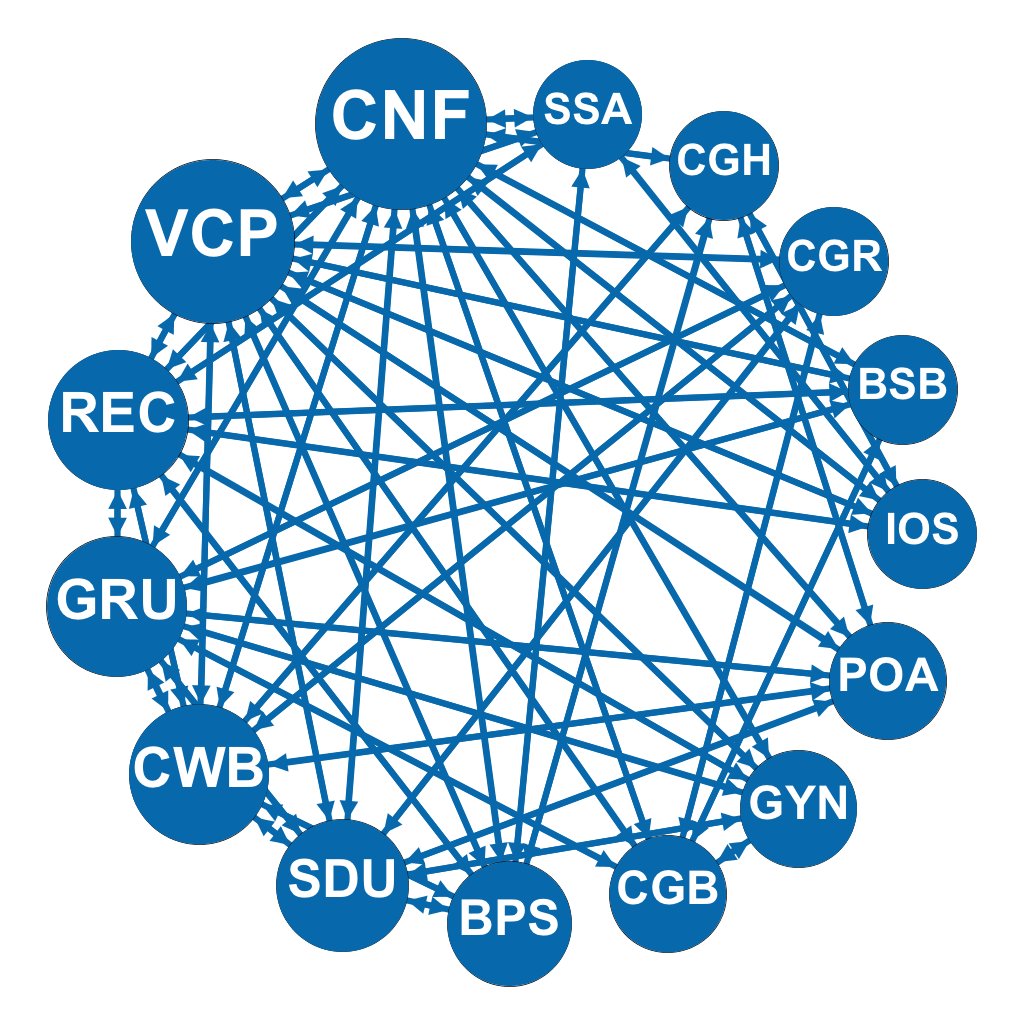}
	     \label{subfig:CDA2}
			  }
	\subfigure[Tam's core on June 2015]{ 
    	\includegraphics[width=0.35\textwidth]{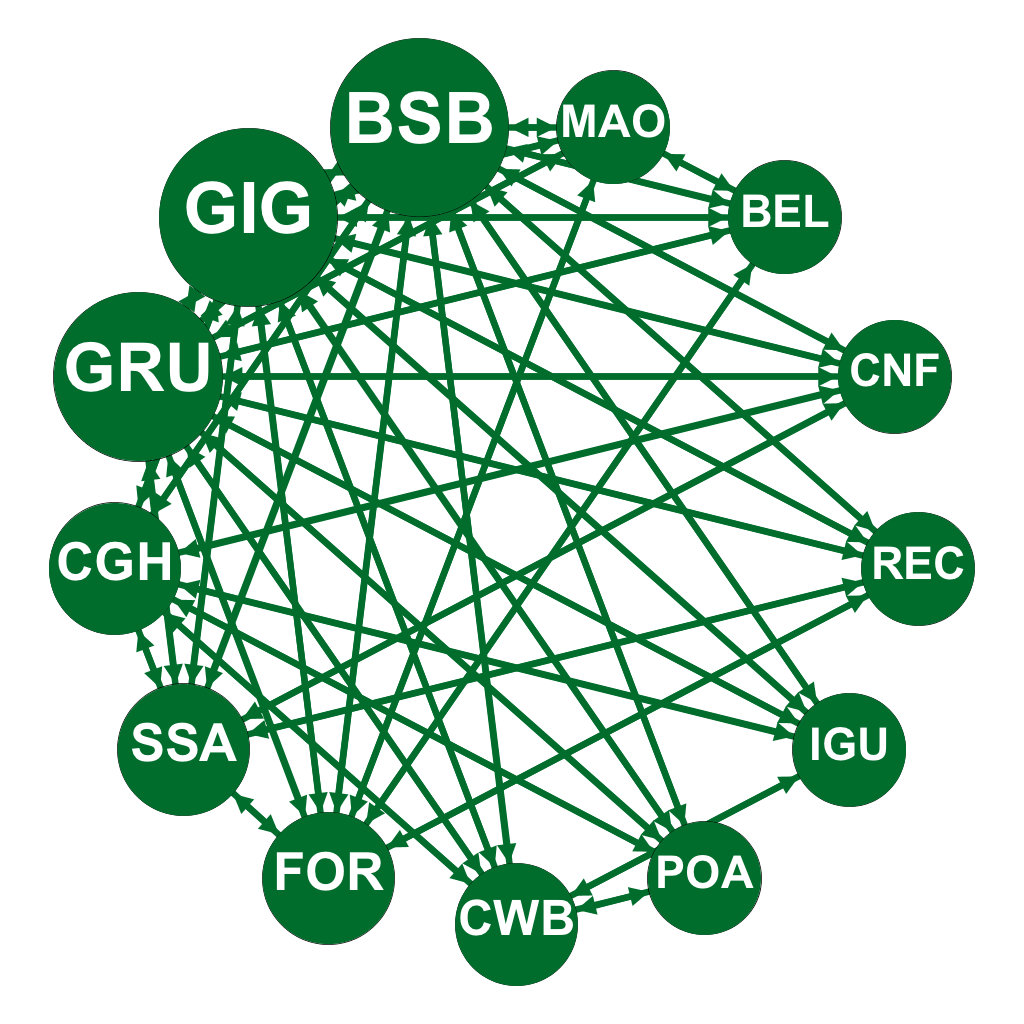}
      	\label{subfig:core-tam-2015}
	   }
	\subfigure[Tam's core on May 2016]{
		\includegraphics[width=0.35\textwidth]{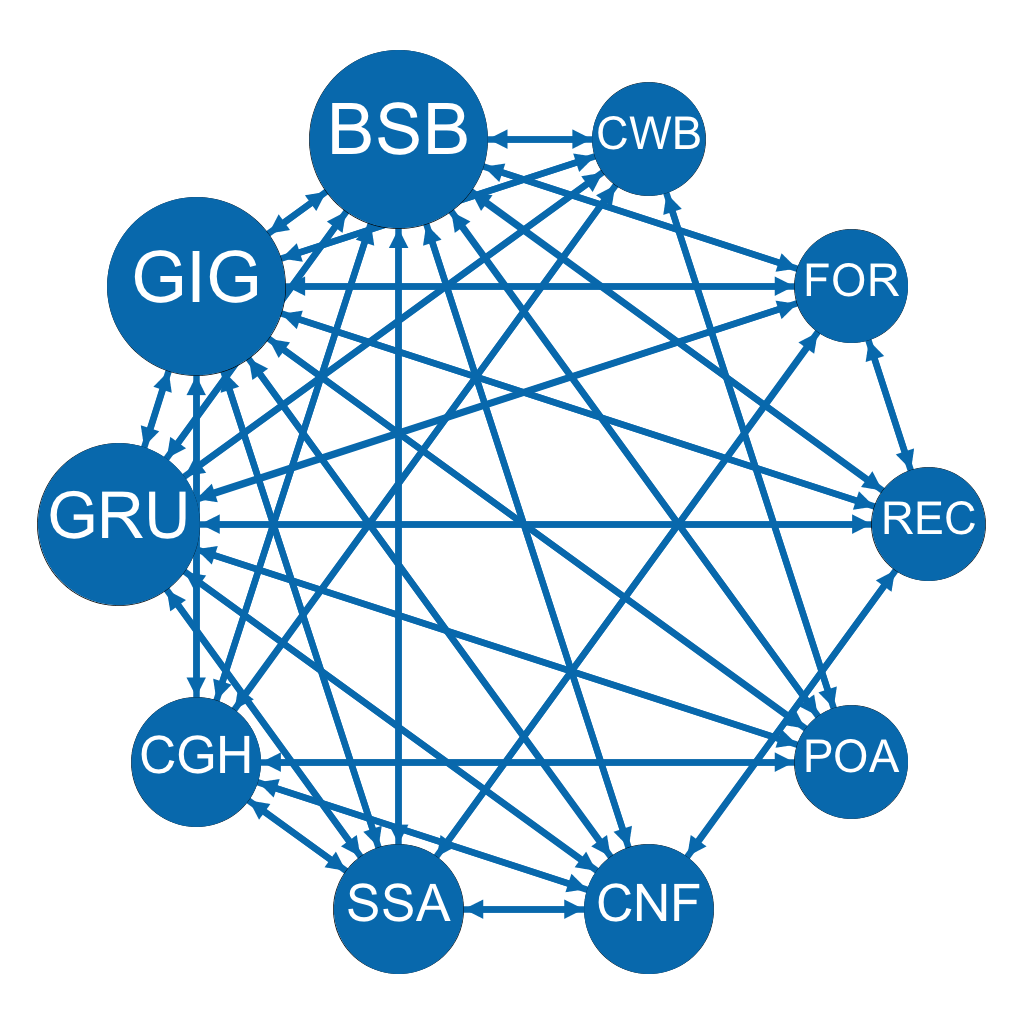}
	     \label{subfig:core-tam-2016}
			  }
	\subfigure[Avianca's core on June 2015]{ 
    	\includegraphics[width=0.35\textwidth]{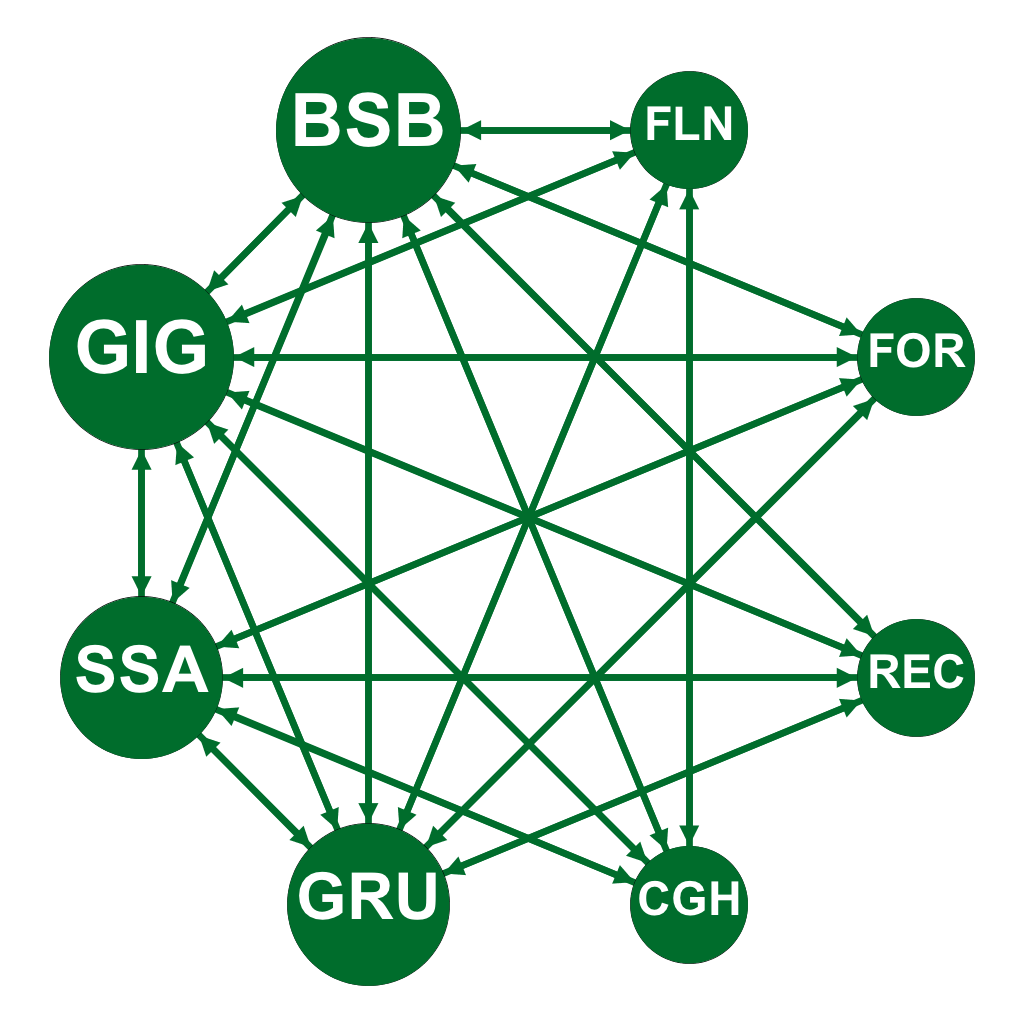}
      	\label{subfig:A2015}
	   }
	\subfigure[Avianca's core on May 2016]{
		\includegraphics[width=0.35\textwidth]{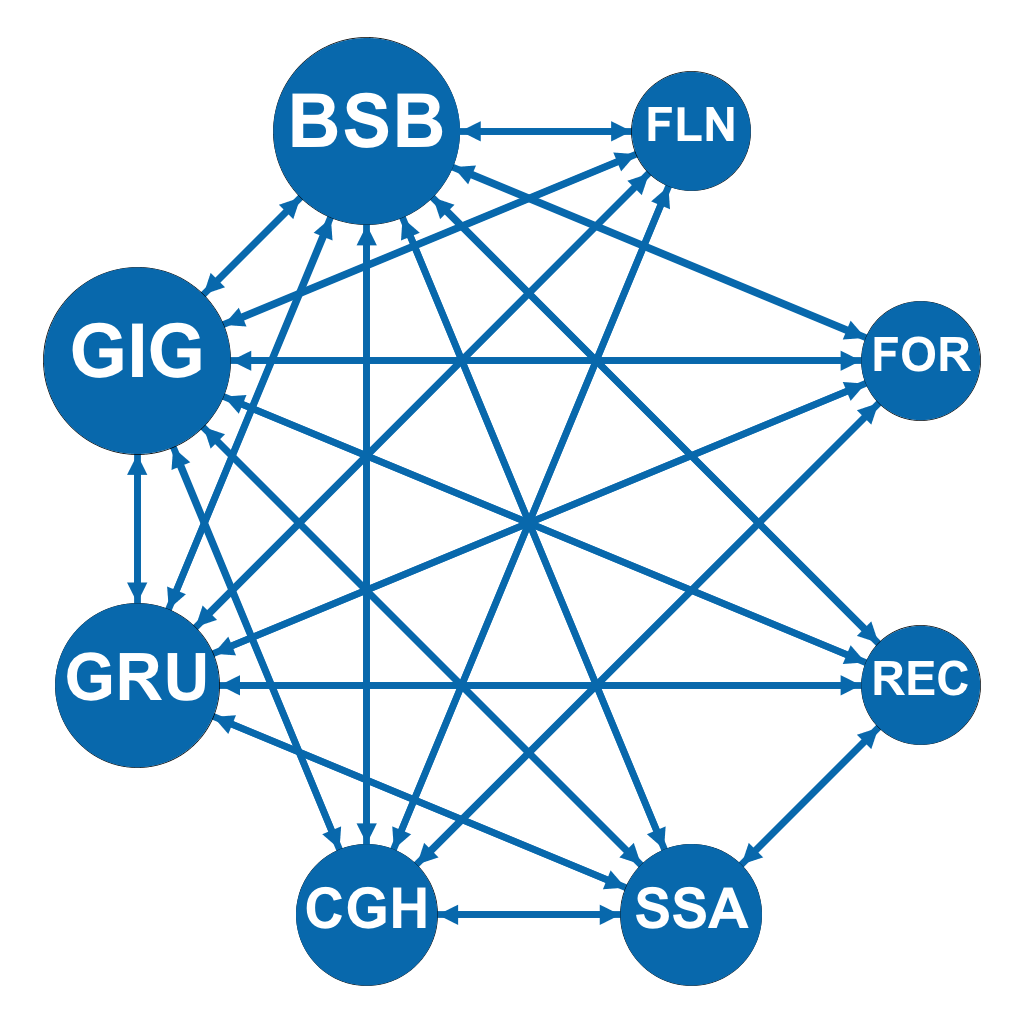}
	     \label{subfig:A2016}
			  }
			  \caption{Comparison of the maximum K-Core for GOL, Azul, TAM, and Avianca in June 2015 and May 2016 using the digraph format, i.e., the route-based perspective.}
			  \label{fig:comp-routes}
\end{figure}

Figure~\ref{fig:comp-routes} shows the maximum K-Cores in a route-based perspective for Gol, Azul, TAM, and Avianca in June 2015 and May 2016. This figure allows the comparison of the main structural core of main airlines in Brazil and how each company adapted to the new scenario imposed by the economical crisis when we compare the cores from 2015 to 2016, as we discuss in further detail next.

Figures~\ref{subfig:gol-core-2015} and~\ref{subfig:gol-core-2016} show how Gol structurally changed its air transportation network between 2015 and 2016. By 2015, Gol served 56 airports in total~(see Table~\ref{tab:num_air}), but with 13 airports in its main core, corresponding to $K = 12$. By 2016, in contrast, Gol served 53 airports, with 8 airports in its main core, which also corresponds to $K = 12$. The reduction on the number of airports in the main core indicates that Gol's strategy was to keep the level of connectivity on the main core of its air transportation network ($K = 12$ in both periods of time), but reduced the number of served airports at this level of connectivity  to accommodate the decreasing demand.

Figures~\ref{subfig:CDA1} and~\ref{subfig:CDA2} illustrate how Azul changed its air transportation network between 2015 and 2016. We clearly see how the strategy of Azul is to have a more distributed main core. Although it has a similar level of connectivity as Gol or TAM, Azul's main core includes a significantly larger number of airports. By 2015, Azul served 100 airports, with 21 airports in its main core with $K = 10$. By 2016, Azul served 94 airports, with 15 airports in its main core, which kept $K = 10$, as well. The adaptation of Azul's strategy between 2015 and 2016 was similar to Gol's: to reduce the number of served airports while keeping the level of connectivity of the main core, even with a reduced core. Anyway, Azul kept having the largest capillarity serving a significant number of smaller airports.
 
Figures~\ref{subfig:core-tam-2015} and~\ref{subfig:core-tam-2016} show how TAM changed its air transportation network between 2015 and 2016. By 2015, TAM served 43 airports, with 13 airports in its main core with $K = 10$. By 2016, TAM served 46 airports, with 10 airports in its main core, which kept the level of connectivity, i.e. $K = 10$. By 2015, TAM presented a level of connectivity in its main core similar to Azul's, but including much less airports with that level of connectivity (a number of airports similar to Gol's core with a higher level of connectivity). By 2016, TAM also adapted its strategy by keeping the same level of connectivity in its main core, but reducing the number of airports at the core, although with less intensity than the adaptation of Gol. 
Unlike Gol, TAM's strategy  was to start serving a few more airports, despite reducing the size of its main core. 

Finally, Figures~\ref{subfig:A2015} and~\ref{subfig:A2016} show the main core of the air transportation network of Avianca in June 2015 and May 2016, respectively. In contrast with the other main airline companies, Avianca kept exactly the same main core with the same level of connectivity between 2015 and 2016, although it has increased the number of routes and flights~(see Tables~\ref{tab:num_rotas} and~\ref{tab:num_voos}). This outcome indicates that Avianca's strategy in face of the crisis was to keep its main core, whereas adding some routes and increasing the number of flights in existing routes in order to take benefit of the reduction in activity of the competing companies to gain market.

\subsection{Multi-digraph analysis: The flight-based perspective}
\label{subsec:multidigraph}

In this section, we analyze the results obtained from the multi-digraph corresponding to the flights in the Brazilian air transportation network. Multi-digraph is the oriented graph that may have multiple~(or parallel) edges, i.e., two vertices may be connected by more than one edge. In other words, a single route between two airports, as seen in the digraph format in Section~\ref{subsec:digraph}, can be split into multiple flights that use that route. This feature of the multi-digraph can thus better represent the real usage of a route, since typically many airport pairs are connected by multiple flights along a single day. The multi-digraph representation thus allows quantifying how busy a route between an airport pair is.
As shown in Section~\ref{subsubsec:sub-mag}, using the sub-MAG operation, with a single MAG, it is possible to separate all the airlines and analyze their characteristics individually. Similarly to the route-based view presented in Section~\ref{subsec:digraph}, in this section a flight-based view is provided using a multi-digraph for each airline company in each of the considered time periods. 

\begin{figure}[ht]
  \begin{center}
  \subfigure[2015]{
  \includegraphics[width=0.47\textwidth]{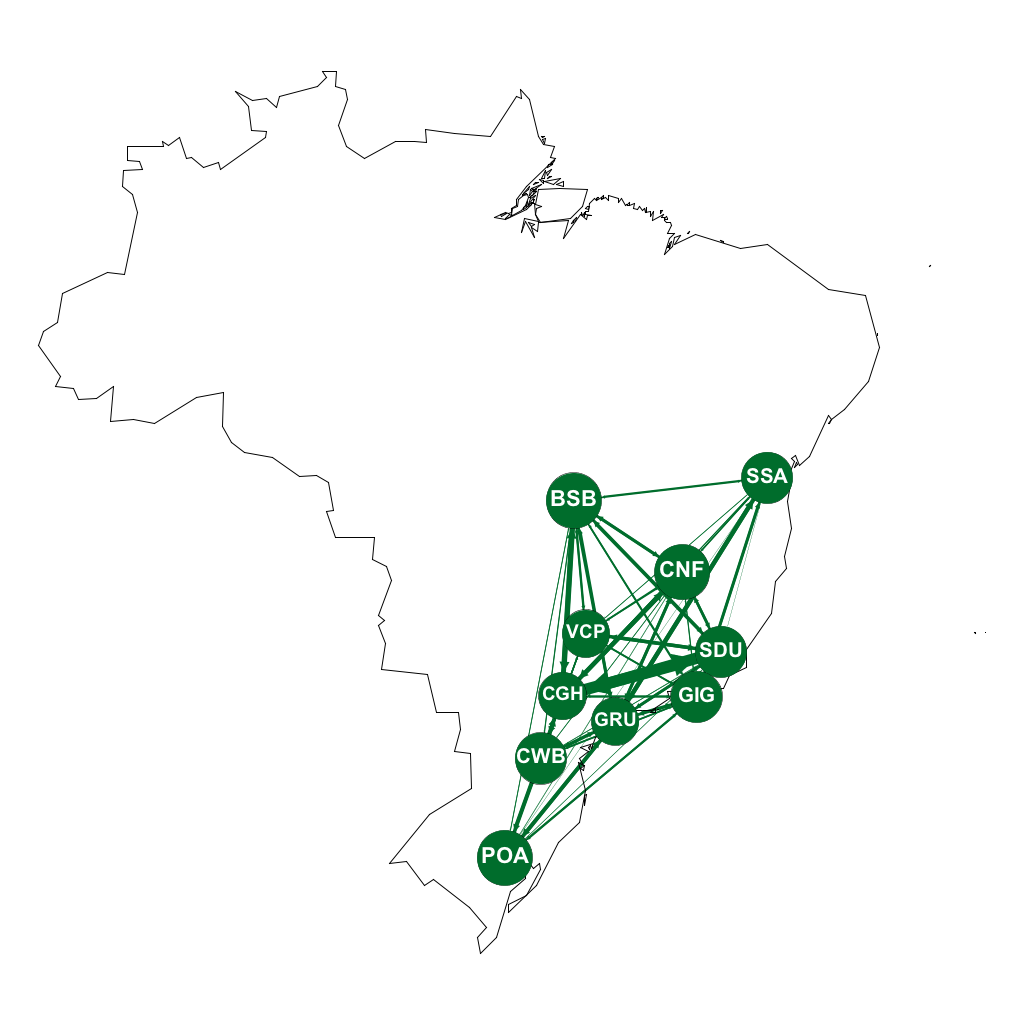}
  \label{fig:multi-core-2015}
  }
  \subfigure[2016]{
  \includegraphics[width=0.47\textwidth]{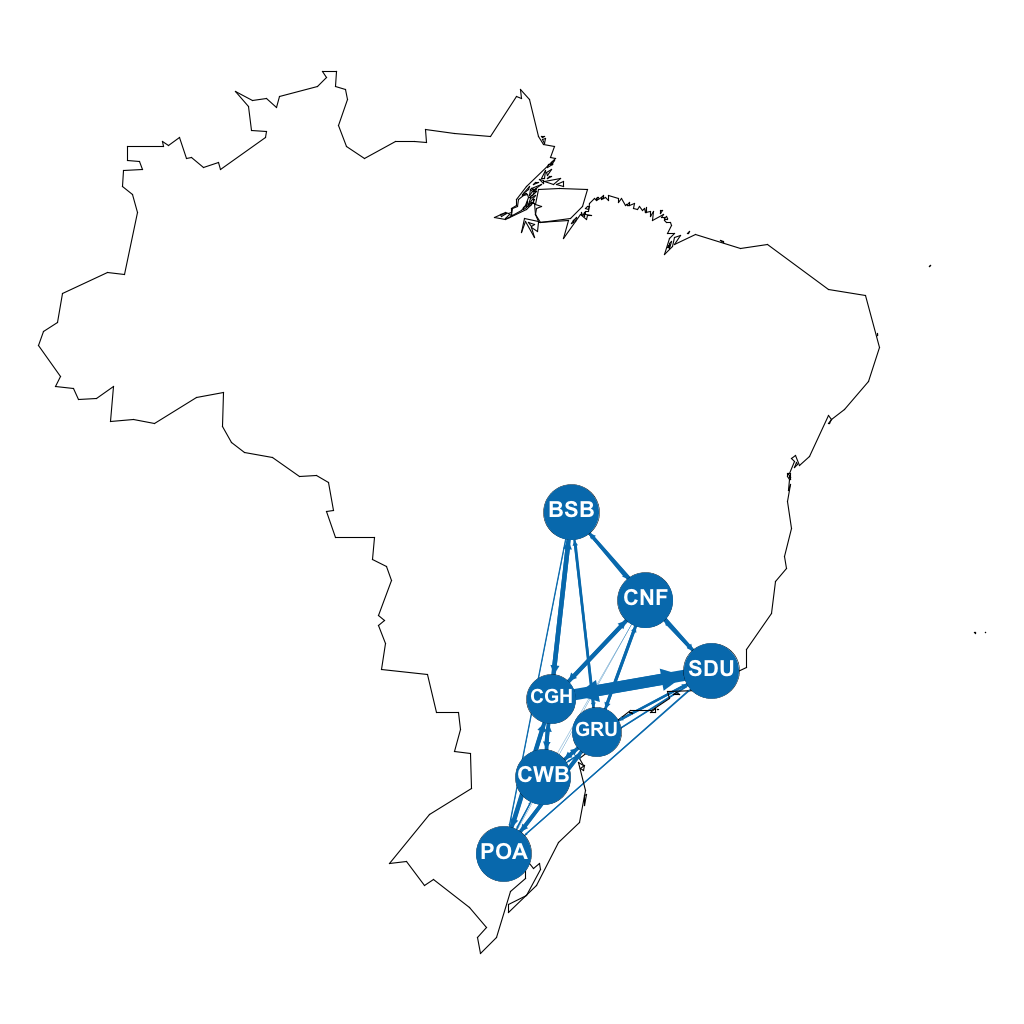}
    \label{fig:multi-core-2016}
  }
  \caption{The geo-referenced K-Core of the Brazilian air transportation network with a multi-digraph (flight-based) perspective: (a) as of July 2015 for the maximum value of K=967 (10 airports and 7475 flights in the maximum K-Core); and (b) as of May 2016 for the maximum K=813 (7 airports and 4130 routes in the maximum K-Core). }
  \label{fig:multi-core}
  \end{center}
\end{figure}

Figure~\ref{fig:multi-core}  shows the geo-referenced K-Core of the Brazilian air transportation net- work with a multi-digraph (flight-based) perspective for 2015 and 2016. The trend to- wards concentration emerges in Figure 3(b) showing the busiest airports in the country.

\begin{table}[ht]  	
 \centering
 \caption{Comparison of maximum K-Cores in the multi-digraph representation.}
 \label{tab:core_mul}
 \vspace{0.5cm}
  \begin{tabular}{|l|r|r|r|r|r|r|}
    \hline
     \multirow{2}{*}{Airlines} &
      \multicolumn{2}{c|}{Core (K)} &
      \multicolumn{2}{c|}{\# Airports} &
      \multicolumn{2}{c|}{\# Flights} \\
      	   & 2015 & 2016 & 2015 & 2016 & 2015 & 2016 \\
   \hline                               
    Gol & 395 & 308 & 2 & 2 & 395 & 308\\
    Azul & 288 & 288 & 6 & 6 & 1131  & 1120\\
    TAM & 381 & 345 & 2 & 2 & 381 & 345 \\
    Avianca & 160 & 160 & 2 & 2 & 160 & 160\\
  \hline 
  \end{tabular}
\end{table}

We analyze the connectivity of the maximum K-Core in the multi-digraph corresponding to flight network of each of the four main airline companies in Brazil. Table~\ref{tab:core_mul} shows the maximum core number~$K$ as well as the number of airports and flights in the maximum K-Core of each of the main airline companies, both in June 2015 and May 2016. The maximum core number~$K$ for Azul and Avianca has remained the same between the two time periods, as well as the number of airports in their maximum K-Core, although Azul has reduced the number of flights in its core network. Avianca, in contrast, kept the exact same K-Core between the two time periods. Meanwhile, Gol and TAM have changed their core structures in such a substantive way that this was reflected in a reduction of the maximum core number~$K$ for their core structures. 

\begin{figure}[p]
\centering
	\subfigure[Gol's core on June 2015]{ 
    	\includegraphics[width=0.35\textwidth]{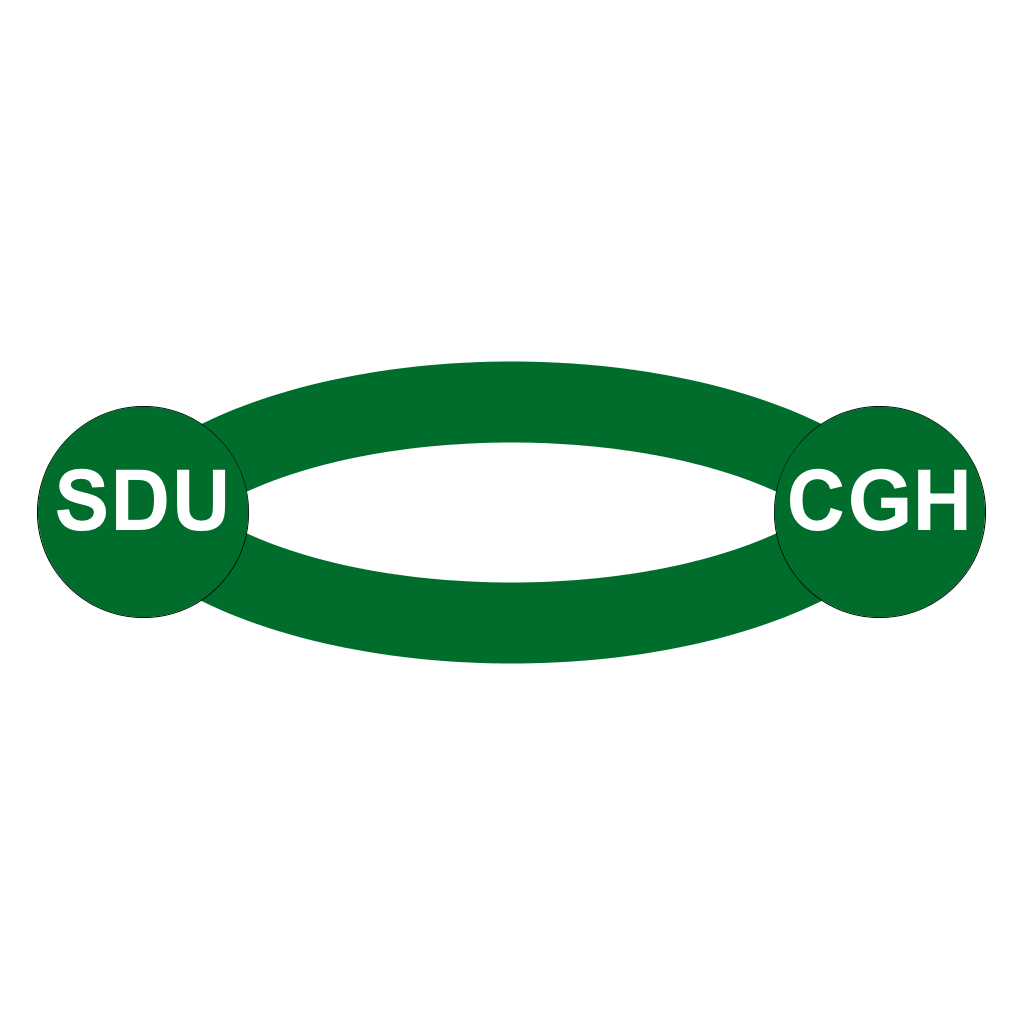}
      	\label{subfig:mul-gol-core-2015}
	   }
	\subfigure[Gol's core on May 2016]{
		\includegraphics[width=0.35\textwidth]{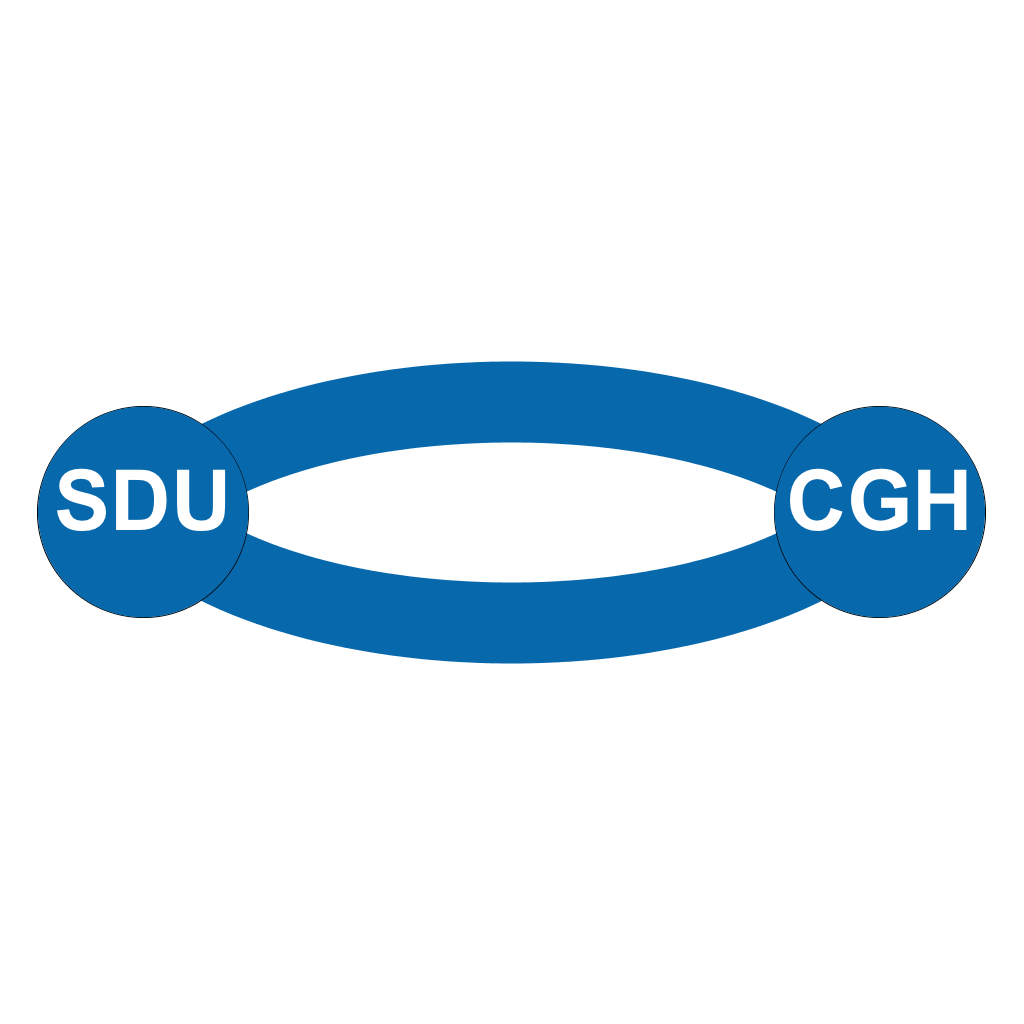}
	     \label{subfig:mul-gol-core-2016}
			  }
	\subfigure[Azul's core on June 2015]{ 
    	\includegraphics[width=0.35\textwidth]{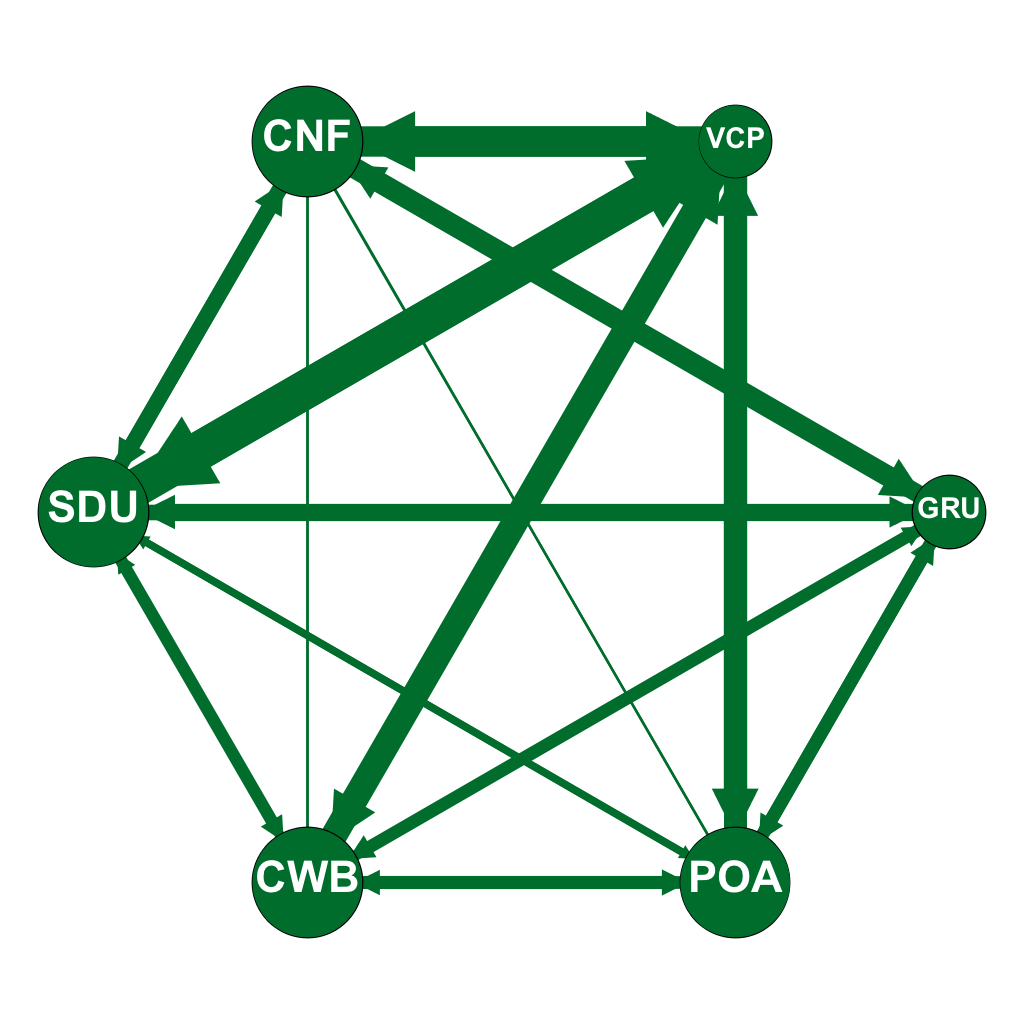}
      	\label{subfig:mul-azul-core-2015}
	   }
	\subfigure[Azul's core on May 2016]{
		\includegraphics[width=0.35\textwidth]{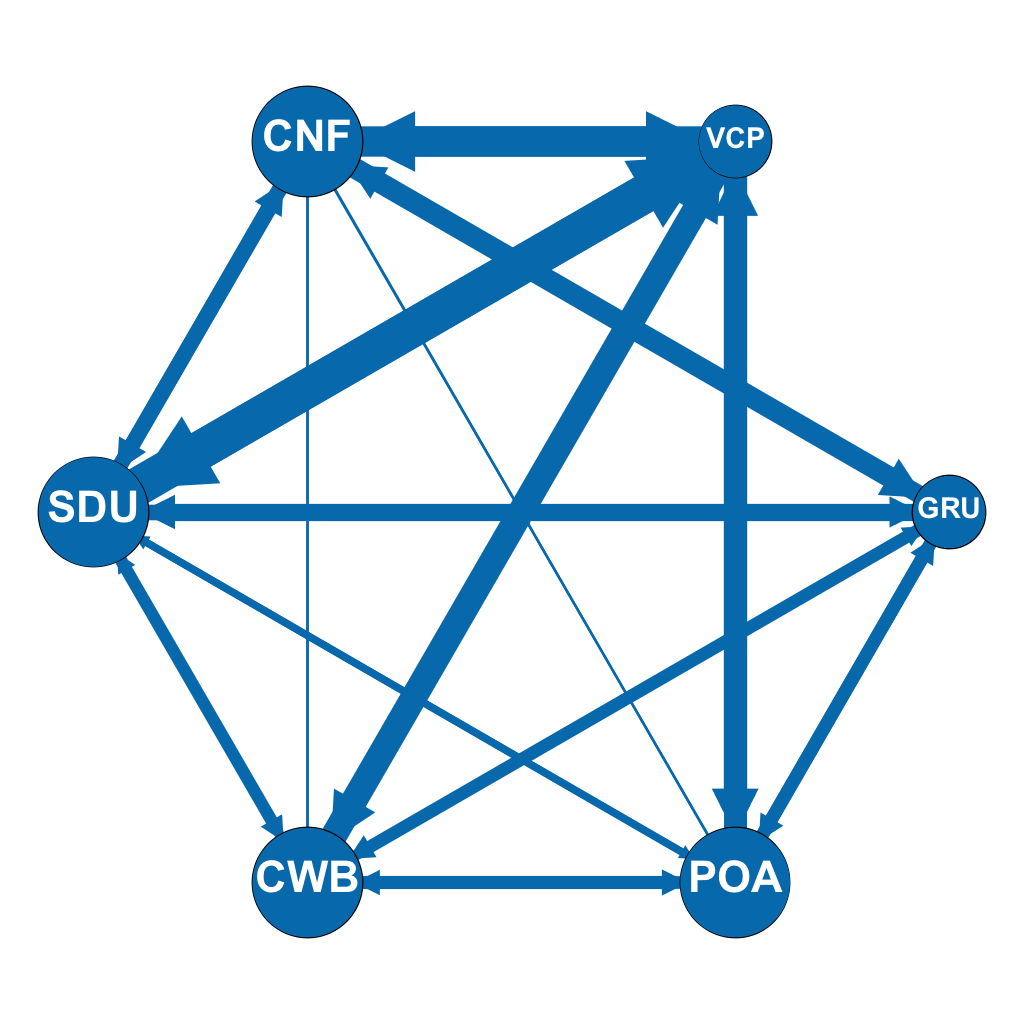}
	     \label{subfig:mul-azul-core-2016}
			  }
	\subfigure[Tam's core on June 2015]{ 
    	\includegraphics[width=0.35\textwidth]{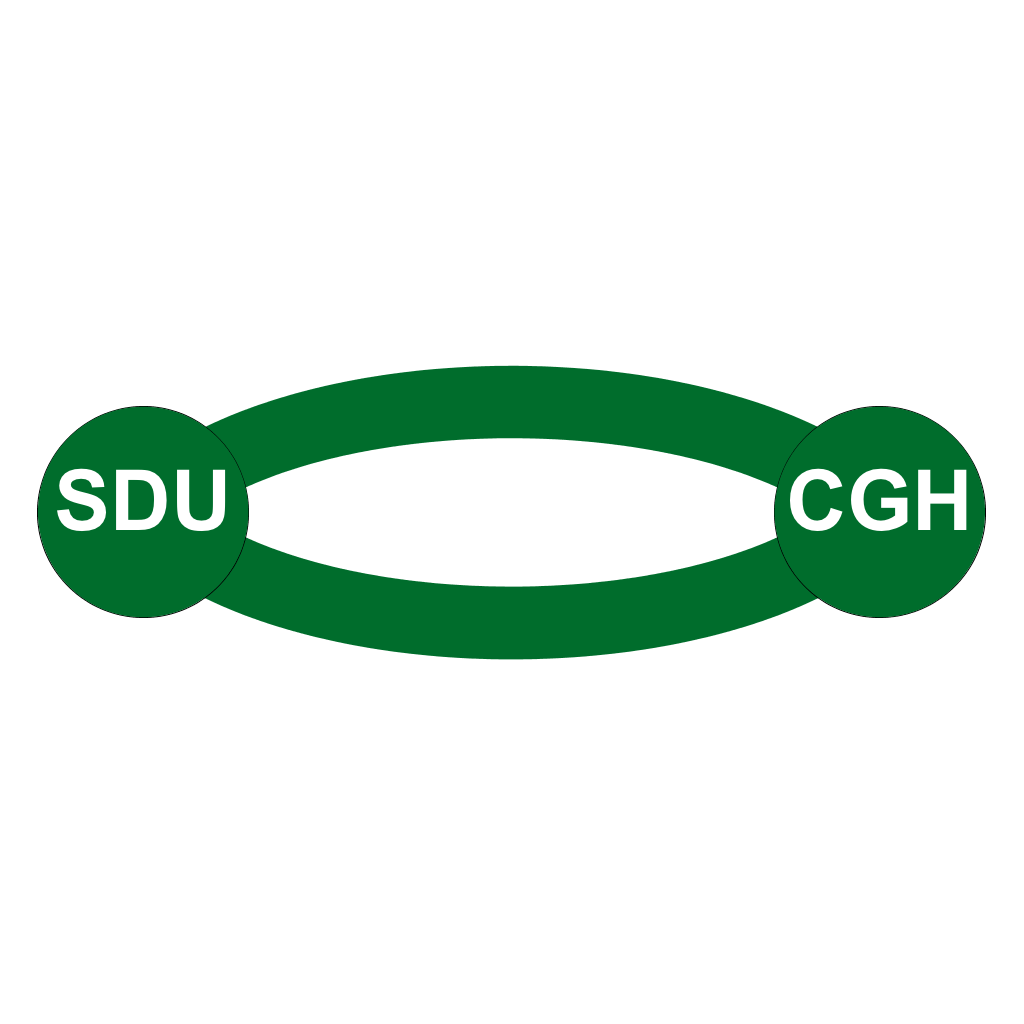}
      	\label{subfig:mul-tam-core-2015}
	   }
	\subfigure[Tam's core on May 2016]{
		\includegraphics[width=0.35\textwidth]{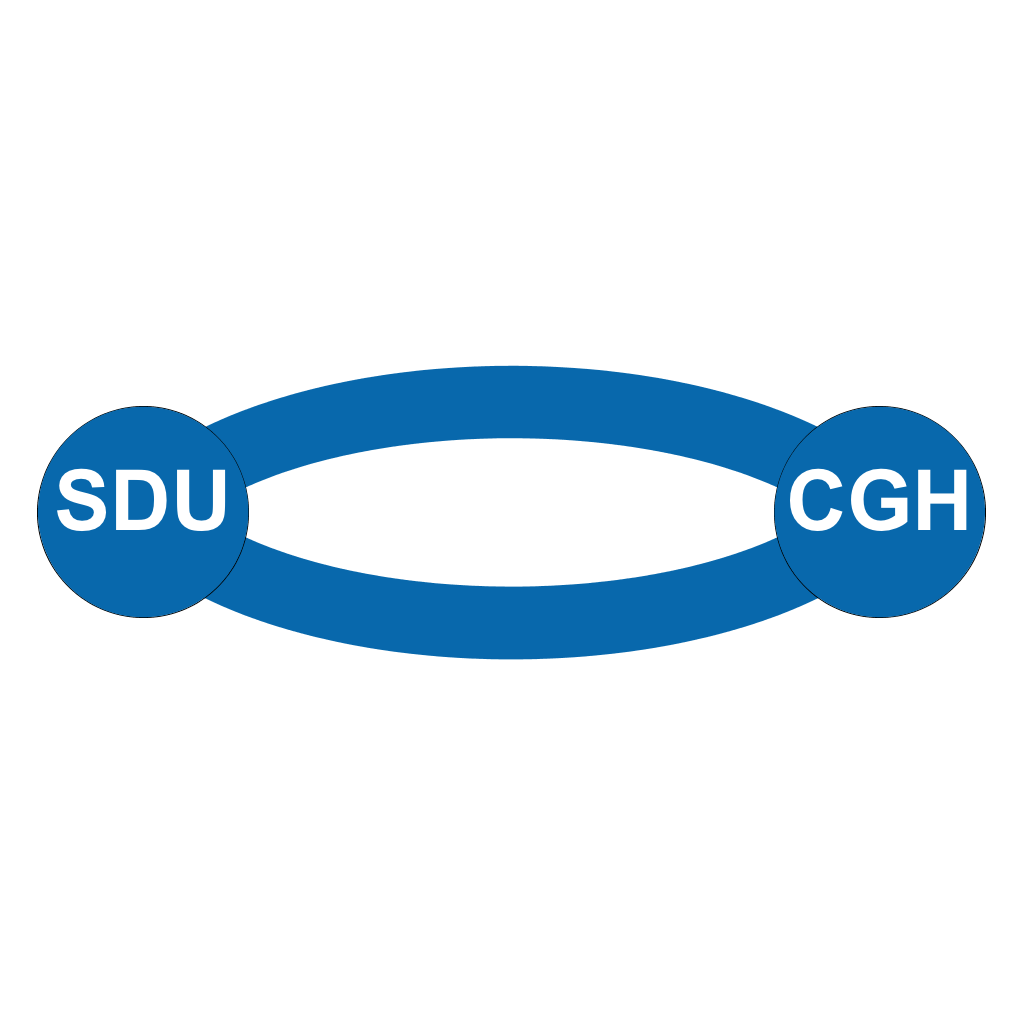}
	     \label{subfig:mul-tam-core-2016}
			  }
	\subfigure[Avianca's core on June 2015]{ 
 		\includegraphics[width=0.35\textwidth]{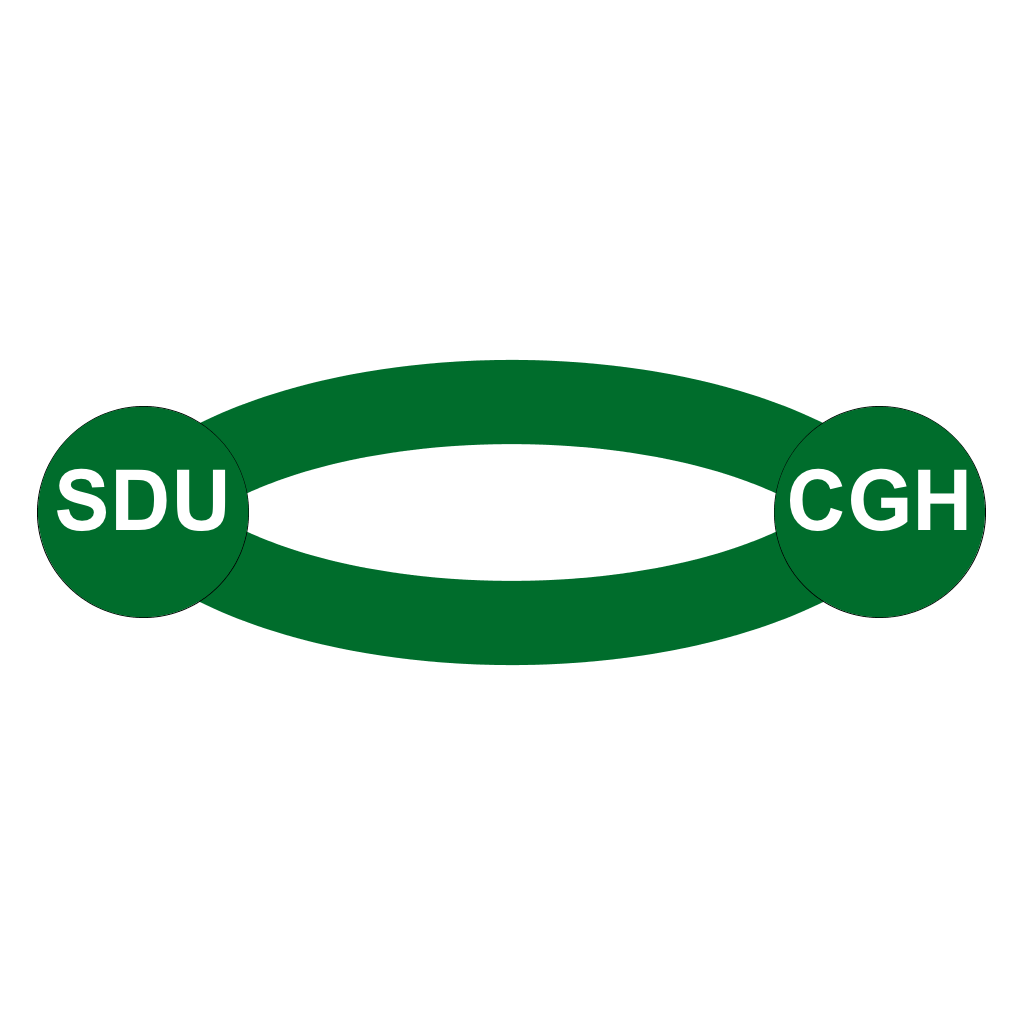}
    	\label{subfig:mul-avianca-core-2015}
	   }
	\subfigure[Avianca's core on May 2016]{
		\includegraphics[width=0.35\textwidth]{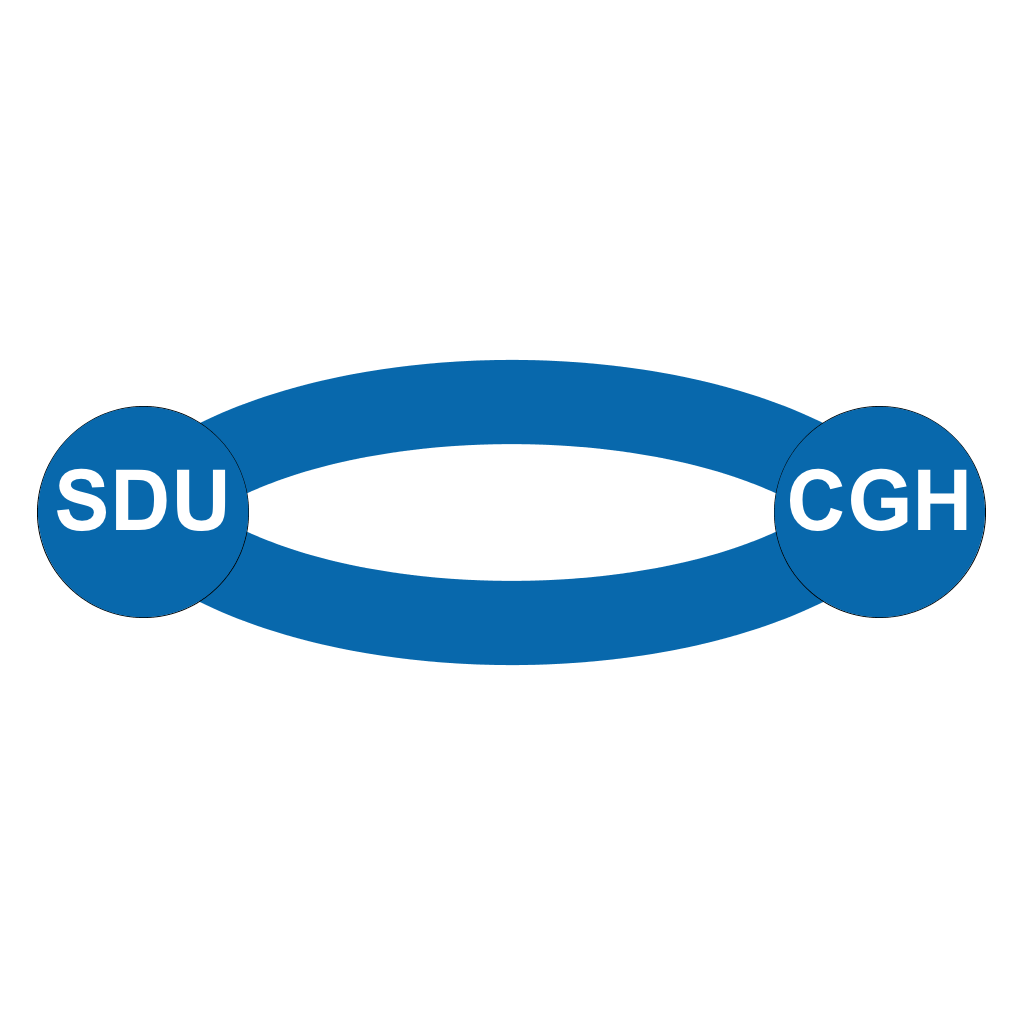}
	     \label{subfig:mul-avianca-core-2016}
			  }
   \caption{Comparison of the maximum K-Core for GOL, Azul, TAM, and Avianca in June 2015 and May 2016 using the multi-digraph format, i.e., the flight-based perspective.}
  \label{fig:comp-flights}
\end{figure}

It is interesting to notice that the small number of airports in the maximum K-Core of each airline in a flight-based perspective clearly indicates the high level of concentration of flights between a very small set of airports. Indeed, Figure~\ref{fig:comp-flights} shows the maximum K-Cores in a flight-based perspective for Gol, Azul, TAM, and Avianca in June 2015 and May 2016. This figure allows the comparison of the main structural core of main airlines in Brazil in a flight-based perspective and how each company adapted to the new scenario imposed by the economical crisis when we compare the cores from 2015 to 2016. Gol, TAM, and Avianca present a minimum configuration of the maximum K-Core composed of a pair of airports, namely the domestic airports Santos Dumont and Congonhas located close to downtown Rio de Janeiro and S\~{a}o Paulo, respectively. These are the two largest cities in the country and this route figures as one of the top~10 busiest domestic passenger airport pairs worldwide. Therefore, it seems natural that most airlines focus a very large number of flights on this particular airport pair as it is responsible for a significant part of their revenue. This concentration is illustrated in Figures~\ref{subfig:mul-gol-core-2015}, \ref{subfig:mul-gol-core-2016},~\ref{subfig:mul-tam-core-2015}, \ref{subfig:mul-tam-core-2016}, \ref{subfig:mul-avianca-core-2015}, and~\ref{subfig:mul-avianca-core-2016}. We also remark that although the core structure for these three airlines has remained concentrated in the Rio de Janeiro -- S\~{a}o Paulo domestic airports pair, the two companies, Gol and TAM, with the largest number of flights in this route have significantly decreased the number of their flights in the core structure with a reduction of 22\% and 9\%, respectively, even though this tends to be one of their most profitable routes. 

Finally, in contrast to the other main airlines in the country, Azul presents a completely different core network from a flight-based perspective. While the other main airlines have, as previously discussed, a flight-based core concentrated in the Rio de Janeiro -- S\~{a}o Paulo domestic airports pair, Azul presents a more well distributed flight-based core composed of 6~airports as shown in Figures~\ref{subfig:mul-azul-core-2015} and \ref{subfig:mul-azul-core-2016}. In a comparison between 2015 and 2016, Azul kept its flight-based core with the same distributed structure with only a slight decrease in the number of flights, a bit less than~1\%. Results thus suggest that Azul follows a fairly different niched strategy as compared to the other main airlines, serving a larger number of airports (thus reaching smaller and medium size cities that the others avoid) and establishing a distributed core of operation. Even in this distributed core, the 
Rio de Janeiro -- S\~{a}o Paulo domestic airports pair is not the most important one, being replaced by the route Rio de Janeiro -- Campinas airports pair (Campinas is relatively close to S\~{a}o Paulo, though) but having other pairs receiving a relatively large number of flights with the Campinas~(VCP) airport acting as a hub in this core. 

\section{Conclusion}
\label{conclusao:}
	
This paper provides a structural analysis of the Brazilian air transportation network in two time periods~(June 2015 and May 2016). A comparative analysis between this two time periods brings results on how the economic crisis impacted the Brazilian air transportation network, as it is shown that the main airline companies have reduced the routes they use between airports and the number of flights that use these routes by 15\% and 20\%, respectively, representing a significative contraction of the Brazilian air transportation network.  

Moreover, adopting MAGs~\cite{Wehmuth2016,Wehmuth2017} for the modeling enabled a multilayer and time-varying structural analysis of the Brazilian air transportation network using a single mathematical object, thus easing the processing and analysis. With such an approach, the multi-layer perspective enabled the unveiling of the particular strategies of each airline to both establish and adapt in a moment of crisis their specific flight networks. Similarly, the time-varying perspective allowed multi-scale analysis considering different time periods, and thus assessing the impact of the economic crisis, but also analyzing the different airlines concerning their routes as well as the flights that use these routes. Altogether, these different perspectives allowed the analysis of the impact due to economic restrictions. Therefore, besides the multilayer and time-varying structural analysis of the Brazilian air transportation network, this paper also acts as a proof-of-concept for the MAG potential for the modeling and analysis of high-order networks. 

Future work includes the development and application of new centrality measures to be applied in MAGs, including time centralities~\cite{Costa2015}, in particular considering applications for the air transportation networks.

   \section*{Acknowledgments}
This work was partially supported by CNPq through the grants authors receive. Authors also particularly acknowledge the INCT in Data Sciences~(CNPq no.~465560/2014-8).

\bibliographystyle{apalike} 
\bibliography{library,local}

\begin{thebibliography}{}

\bibitem[Costa et~al., 2015]{Costa2015}
Costa, E.~C., Vieira, A.~B., Wehmuth, K., Ziviani, A., and da~Silva, A. P.~C.
  (2015).
\newblock Time centrality in dynamic complex networks.
\newblock {\em Advances in Complex Systems}, 18(07n08).

\bibitem[Couto et~al., 2015]{Couto2015}
Couto, G.~S., da~Silva, A. P.~C., Ruiz, L.~B., and Benevenuto, F. (2015).
\newblock Structural properties of the brazilian air transportation network.
\newblock {\em {Anais da Academia Brasileira de Ciências}}, 87:1653--1674.

\bibitem[Holme and Saram{\"a}ki, 2012]{Holme2012}
Holme, P. and Saram{\"a}ki, J. (2012).
\newblock Temporal networks.
\newblock {\em Physics reports}, 519(3):97--125.

\bibitem[Kivel{\"a} et~al., 2014]{Kivela2014}
Kivel{\"a}, M., Arenas, A., Barthelemy, M., Gleeson, J.~P., Moreno, Y., and
  Porter, M.~A. (2014).
\newblock Multilayer networks.
\newblock {\em Journal of Complex Networks}, 2(3):203--271.

\bibitem[Seidman, 1983]{SEIDMAN1983}
Seidman, S.~B. (1983).
\newblock {Network structure and minimum degree}.
\newblock {\em Social Networks}, 5(3):269--287.

\bibitem[Sternberg et~al., 2016]{Sternberg2016}
Sternberg, A., Carvalho, D., Murta, L., Soares, J., and Ogasawara, E. (2016).
\newblock An analysis of {B}razilian flight delays based on frequent patterns.
\newblock {\em Transportation Research Part E: Logistics and Transportation
  Review}, 95:282--298.

\bibitem[Tsiotas and Polyzos, 2015]{Tsiotas2015}
Tsiotas, D. and Polyzos, S. (2015).
\newblock Decomposing multilayer transportation networks using complex network
  analysis: a case study for the {G}reek aviation network.
\newblock {\em Journal of Complex Networks}, 3(4):642--670.

\bibitem[Verma et~al., 2014]{Verma2014}
Verma, T., Ara{\'{u}}jo, N. A.~M., and Herrmann, H.~J. (2014).
\newblock {Revealing the structure of the world airline network}.
\newblock {\em Scientific Reports}, 4:1--6.

\bibitem[Wehmuth et~al., 2016]{Wehmuth2016}
Wehmuth, K., Fleury, {\'{E}}., and Ziviani, A. (2016).
\newblock {On MultiAspect graphs}.
\newblock {\em Theoretical Computer Science}, 651:50--61.

\bibitem[Wehmuth et~al., 2017]{Wehmuth2017}
Wehmuth, K., Fleury, {\'{E}}., and Ziviani, A. (2017).
\newblock {MultiAspect Graphs: Algebraic Representation and Algorithms}.
\newblock {\em Algorithms}, 10(1):1--36.

\bibitem[Wehmuth et~al., 2015]{Wehmuth2015-dsaa}
Wehmuth, K., Ziviani, A., and Fleury, E. (2015).
\newblock A unifying model for representing time-varying graphs.
\newblock In {\em Proc. of the IEEE Int. Conf. on Data Science and Advanced
  Analytics (DSAA)}, pages 1--10.

\bibitem[Wei et~al., 2014]{Wei2014}
Wei, P., Chen, L., and Sun, D. (2014).
\newblock {Algebraic connectivity maximization of an air transportation
  network: The flight routes’ addition/deletion problem}.
\newblock {\em Transportation Research Part E: Logistics and Transportation
  Review}, 61:13--27.

\end{thebibliography}

\end{document}